\documentclass[preprint2]{aastex}
\input psfig

\shorttitle{The HET M-Dwarf Planet Search}
\shortauthors{Endl M., Cochran W.D., et al.}
\begin{document}

\title{A Dedicated M-Dwarf Planet Search Using The \\ Hobby-Eberly Telescope
\footnote{Based on data collected with the Hobby-Eberly Telescope, which is operated
by McDonald Observatory on behalf of The
     University of Texas at Austin, the Pennsylvania State University, Stanford University,
     Ludwig-Maximilians-Universit\"at M\"unchen, and Georg-August-Universit\"at G\"ottingen.}}

\author{Michael Endl ,William D. Cochran, Robert G. Tull \and Phillip J. MacQueen}

\affil{McDonald Observatory, The University of Texas at Austin, Austin, TX 78712, USA}

\email{mike@astro.as.utexas.edu; wdc@astro.as.utexas.edu; rgt@astro.as.utexas.edu; pjm@astro.as.utexas.edu}

\begin{abstract}
We present first results of our planet search program using the 9.2 meter Hobby-Eberly Telescope (HET) at McDonald
Observatory to detect planets around M-type dwarf stars via high-precision radial velocity (RV)
measurements. Although more than $100$ extrasolar planets have been found around solar-type
stars of spectral type F to K, there is only a single M-dwarf (GJ~876, Delfosse et al.~1998;
Marcy et al.~1998; Marcy et al.~2001) known to harbor a planetary system. 
With the current incompleteness of
Doppler surveys with respect to M-dwarfs, it is not yet possible to decide whether this is due to a fundamental
difference in the formation history and overall frequency of planetary systems in the low-mass regime of the
Hertzsprung-Russell diagram, or simply an observational bias. Our HET M-dwarf survey plans to survey
100 M-dwarfs in the next 3 to 4 years with the primary goal to answer this question. 
Here we present the results from the first year of the survey which show that
our routine RV-precision for M-dwarfs is $6~{\rm m\,s}^{-1}$. We found that GJ~864 and GJ~913 
are binary systems with yet undetermined periods, while $5$ out of $39$ M-dwarfs reveal a high
RV-scatter and represent candidates for having short-periodic planetary companions. 
For one of them, GJ~436 (rms = $20.6~{\rm m\,s}^{-1}$), we have already obtained follow-up observations
but no periodic signal is present in the RV-data.
      
\end{abstract}


\section{Introduction}

M-dwarf stars form the majority of stars in our galaxy, both in numbers as well as in total mass.
However, Doppler surveys looking for stellar reflex motions due to 
planetary companions via precise radial velocity measurements, have traditionally focused on the 
brighter F,G and K-type stars in order to obtain a sufficient signal-to-noise ratio of the high resolution
(typically $R > 50,000$) spectra. 
This has led to the discovery of more than 100 extrasolar giant planets orbiting
solar-type stars (Mayor \& Queloz 1995, see e.g. Fischer et al. 2003 for recent detections). 
Due to their intrinsic faintness and thus need of large
aperture telescopes, M dwarfs constitute only a small fraction of the target-samples of these
surveys. Up to now, there is only a single M-dwarf, GJ~876 (M4V, M=$0.3~{\rm M}_{\odot}$), known to possess a 
planet with a minimum mass of $m\sin i \approx 2~{\rm M}_{\rm Jup}$ in a $60$ day orbit (Delfosse 
et al.~1998; Marcy et al.~1998). Follow-up observations 
revealed the presence of a second companion with $m\sin i = 0.56~{\rm M}_{\rm Jup}$ 
moving in a 2:1 mean motion resonance (Marcy et al.~2001). 
In combination with the RV data Benedict et al.~\cite{fritz1} used the HST Fine Guidance Sensors (FGS)
to measure the
astrometric perturbation of GJ~876 caused by the outer planet, and
thus to determine the orbital inclination and the true mass of the companion.  

There were few attempts to target specifically
M-dwarfs to look for substellar companions.
Starting in 1995 the Delfosse et al.~\cite{delfosse2} Doppler survey monitors a volume-limited 
sample of $127$ M-dwarfs 
with the ELODIE spectrograph at Observatoire de Haute-Provence and the CORALIE instrument at ESO La Silla. 
Their long-term RV-precision spans from $10~{\rm m\,s}^{-1}$ for the brighter stars to $70~{\rm m\,s}^{-1}$ for
the fainter objects. Beside the planetary companion to GJ~876, they also discovered several spectroscopic
stellar binary systems in their sample (Delfosse et al.~1999).
The Keck Doppler survey of the Berkeley group includes $\approx 150$ M-stars and they achieve
a long-term RV precision of $3 - 4~{\rm m\,s}^{-1}$ (Vogt et al.~2000).   
The Keck Hyades program of Cochran et al.~\cite{bill2002} also contains a subsample of $20$ M-stars. 
In the southern hemisphere 
we have been using the ESO VLT and the UVES spectrograph to observe $25$ M-stars for $2$ years now. 
The high RV precision and sampling density we obtain for Proxima Cen (M5V) would
have already allowed us to detect very low-mass planets ($m\sin i = 0.0126 - 0.019~{\rm M}_{\rm Jup}$ or
$m\sin i = 4 - 6~{\rm M}_{\oplus}$) inside
the habitable zone of that star (Endl et al.~2003), while in the case of
Barnard's star (M4V) we successfully measured the secular acceleration of the RV 
for the first time and 
discovered a correlation between our RV data and the H$\alpha$ emission level for this M-dwarf (K\"urster 
et al.~2003). 
   
With the exception of the Delfosse et al. and our VLT program, none of the above mentioned surveys 
aim for M-dwarfs as primary targets and in general low-mass stars constitute only a subsample of
larger targetlists. Our HET program only observes M-dwarfs, since we are especially interested
in the prevailance of planetary systems in this section of the HR-diagram. 
 
\section{Planets around M-dwarfs}

M-dwarfs form the majority of stars in our galaxy, both in
total numbers as well as in cumulative mass. Knowledge about the properties and 
frequency of planetary systems for this type of stars would therefore have 
vast implications on the overall statistics of planetary companions in the galaxy.  

So far no detailed physical models for planet formation in general or as a function
of stellar mass in particular are available.
To our knowledge, Nakano~\cite{nakano} was the first to study planet formation timescales 
for stars of varying mass and lifetimes.  
Wetherill~\cite{wetherill} simulated the formation of terrestrial planets for stars 
of different mass and disks of different surface densities, and found that 
the end-products of planetesimal accretion are rather insensitive to the stellar mass. 
However, the low-mass star 
chosen in both studies has a mass of $0.5~{\rm M}_{\oplus}$, which is the high end of the
M-dwarf mass distribution. Since formation of gas giant planets is even less understood (it
is not clear yet whether jovian planets form in a ``bottom-up'' process, i.e. by gas accretion onto
a rocky core of $\approx 10~{\rm M}_{\oplus}$ (e.g. Pollack et al.~1996), or by a type of 
gravitational disk instabilities leading to a ``top-down'' collapse and rapid formation of the
gas giant without the need of a rocky core (e.g. Boss~1997), reliable theoratical predictions 
on the frequency of planetary systems depending on initial conditions like stellar primary and
disk mass as well as disk composition are difficult to make.

Therefore, our goal is to obtain a useful statistical overview of planetary systems in the M-dwarf regime
by observational means. In the near future we will know if GJ~876 remains a special case or if more planets 
around M-dwarfs will be detected by our survey or one of the other programs.

\section{The HET survey and radial velocity results}

Table 1 lists $35$ targets (out of a total of $39$, $4$ candidate objects for short-periodic companions
are not listed) of our survey, the total rms-scatter of their RV-data and the 
duration of monitoring. With a few exceptions the short time span of observations limits our
detection capability to short-periodic planets with periods of a few days. Extension of the
time baseline will allow us to become sensitive also to longer period planets at larger
orbital separations.  

All targets were selected based on the Gliese catalogue of nearby stars (Gliese \& Jahrei\ss~1991)
and the Hipparcos 
astrometric database~\cite{esa}. In order to minimize the effects of stellar activity
on our RV measurements (especially rotational modulation by star spots) we choose only targets 
with no detected or low X-ray emission using the
ROSAT all-sky survey results from H\"unsch et al.~\cite{huensch}.  

In our observing strategy we usually first take $5$ RV measurements per target over a short time to test
for short-periodic variability. If variablity is detected the same target will be re-scheduled for
more intense monitoring. Seemingly constant stars will be re-observed later in our survey and with
a longer cadence in order to check for long-term variability (companions at larger orbital separations).   

This work presents the first results our program, which we plan to continue until we have
sampled $\approx100$ M-dwarfs with the necessary RV-precision to find orbiting giant and lower
mass giant planets. 

For the observations we use the HET High-Resolution-Spectrograph (HRS; Tull 1998) which is
located in an insulated chamber below the telescope. An optical fiber connects the instrument to the
prime focus of the HET and also provides excellent scrambling of the light beam in order to
isolate the spectrograph from variations in the telescope pupil illumination.
All spectra of our program are taken at a resolving power of $R=60,000$ and
an I$_2$ cell is inserted into the light path for self-calibration purposes.

\subsection{RV-precision}

To extract the Doppler shift information from the M-dwarf spectra with the superimposed iodine vapor (I$_2$) 
reference spectrum we perform a full data modeling including the reconstruction of asymmetries of the instrumental 
profile (IP) at the time of observation. In this modeling scheme we subdivide the spectrum into smaller
segments and establish a best-fit model, based on a stellar template spectrum of the target and a 
laboratory spectrum (a Fourier Transform Spectrometer scan) of the I$_2$ cell, for each segment. The
RV of the target is measured differentially to the template while the dense forest of superimposed I$_2$ 
lines delivers the wavelength calibration and the information to reconstruct the IP. The uncertainty 
of the RV measurement is determined from the overall scatter of all segments along a spectrum. A detailed
description of this technique can be found in Butler et al.~(1996) and Endl, K\"urster, Els~(2000). 

\begin{figure} 
\centering{
  \vbox{\psfig{figure=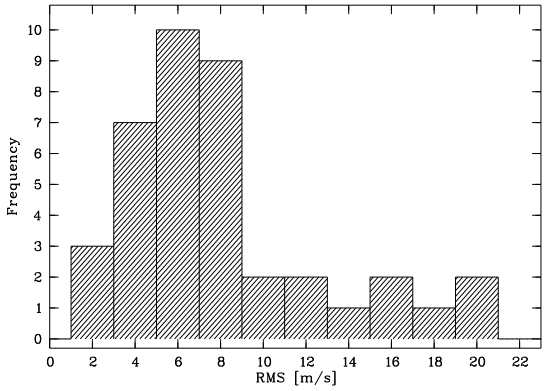,width=8.5cm,height=6.0cm,angle=270}}
   \par
        }
\caption[]{Histogram of the HET/HRS RV-results for $39$ M-dwarfs from our sample.
        The left-hand side contains constant stars and targets which show no excess residual
        signal after subtraction of RV-trends. Fitting a gaussian to this part of the histogram
        yields a mean RV-scatter of $5.7~{\rm m\,s}^{-1}$ and $\sigma = 2.2~{\rm m\,s}^{-1}$.
        The right-hand ``tail'' of the distribution consists of more variable
        stars with rms-values higher than $10~{\rm m\,s}^{-1}$.
        }
 \label{histo}
\end{figure}

Fig.~\ref{histo} shows the histogram of our RV-results.
The bulk of the distribution on the left side consists of the M-dwarfs, which are either constant
within our measurement precision or their residuals after subtraction of RV-trends
do not reveal additional scatter (GJ~623, GJ~708, GJ~864 and GJ~913).   
Fitting a Gaussian to this part of the histogram yields a mean value of $5.7~{\rm m\,s}^{-1}$ with
$\sigma = 2.2~{\rm m\,s}^{-1}$, which is our present mid-term RV-precision for stars
in the magnitude range of $V = 9 - 11$ mag (with the bright ($V=7.5$) M2V star GJ~411 being the sole
exception).
The right-hand ``tail'' of the RV-scatter distribution contains low-amplitude
variable stars (all with rms $> 10~{\rm m\,s}^{-1}$), 
which we want to sample more frequently to check for short-periodic 
planetary companions as possible cause of the higher RV-scatter. 

There are several sources which contribute to the RV-distribution seen in Fig.~\ref{histo}: 
1. intrinsic variability of the targets (i.e. activity induced ``RV-jitter'' or Keplerian
motion due to unknown substellar or stellar companions), 2. the photon noise, 3. the   
stability of the spectrograph used for the RV measurements, and 4. ``algorithmic noise'' introduced
during the data modeling process (e.g. inadequate characterization of the instrumental profile and 
imperfect stellar template deconvolution).
For the M-dwarf HET sample we conclude that photon noise is the major contributor to the RV-scatter.
For earlier type stars of visual magnitude of $V\approx7$~mag, using the same intrumental setup and
analysis technique, we obtain a routine long-term RV-precision of 
$\approx 3{~\rm m\,s}^{-1}$ (Cochran et al.~2003), a factor $2$ better than for the fainter
M-dwarfs.   

\subsection{RV-results}

\begin{figure}
\centering{
        \vbox{\psfig{figure=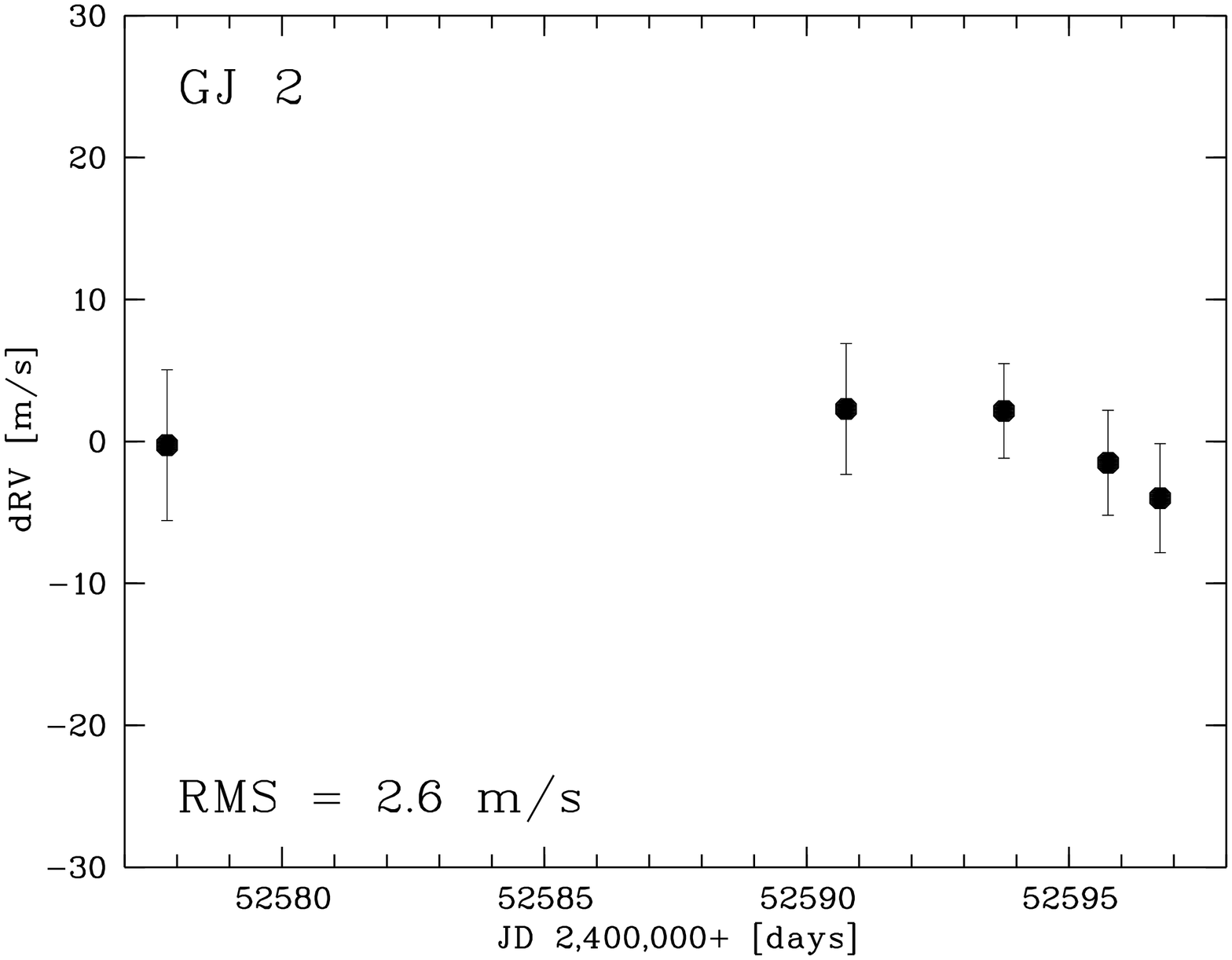,width=9.0cm,height=5.1cm,angle=270}}
        \vbox{\psfig{figure=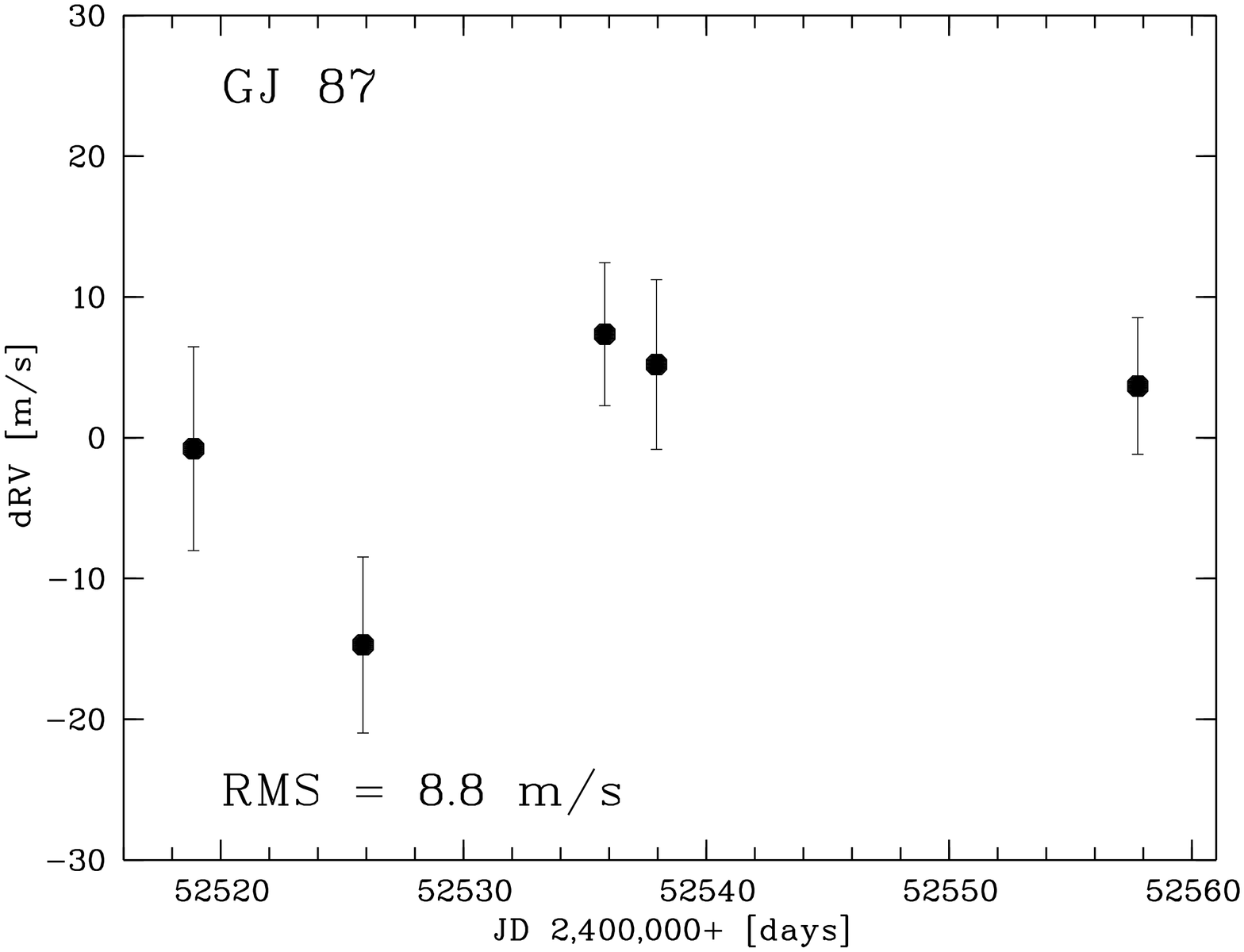,width=9.0cm,height=5.1cm,angle=270}}
        \vbox{\psfig{figure=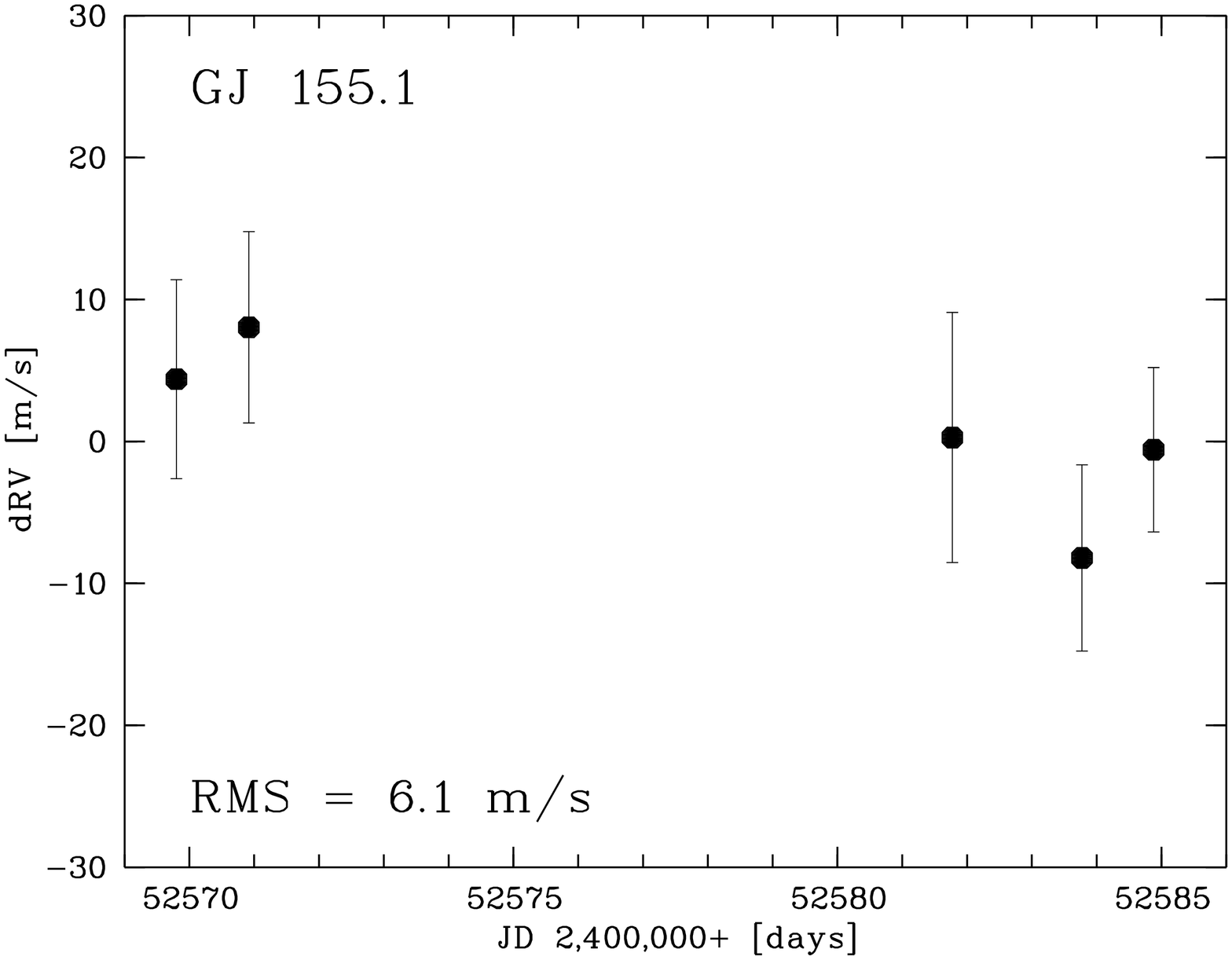,width=9.0cm,height=5.1cm,angle=270}}
        \vbox{\psfig{figure=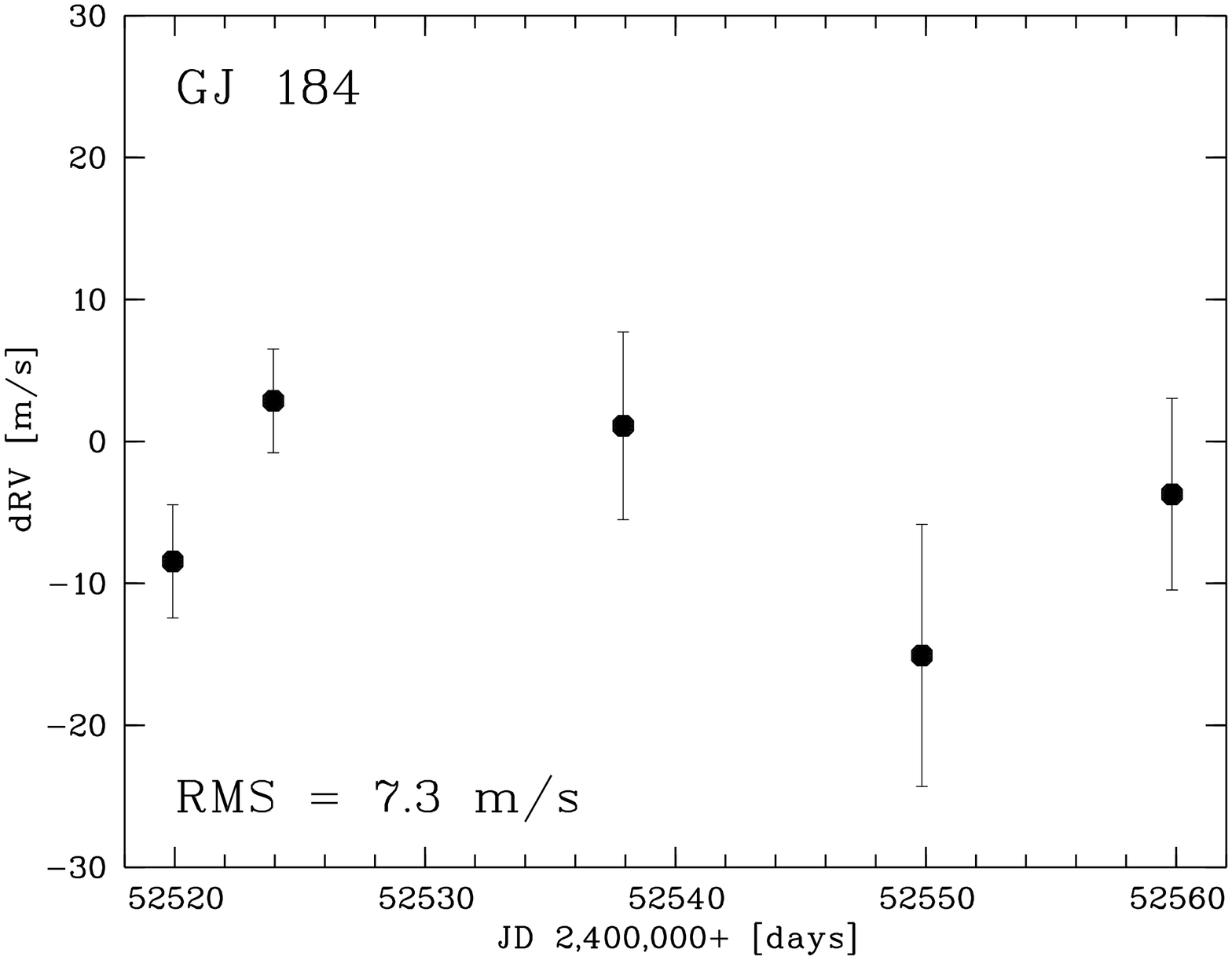,width=9.0cm,height=5.1cm,angle=270}}
   \par
        }
\caption[]{RV results for GJ~2, GJ~87, GJ~155.1 and GJ~184.}
\label{rvsfig1}
\end{figure}

\begin{figure}
\centering{
        \vbox{\psfig{figure=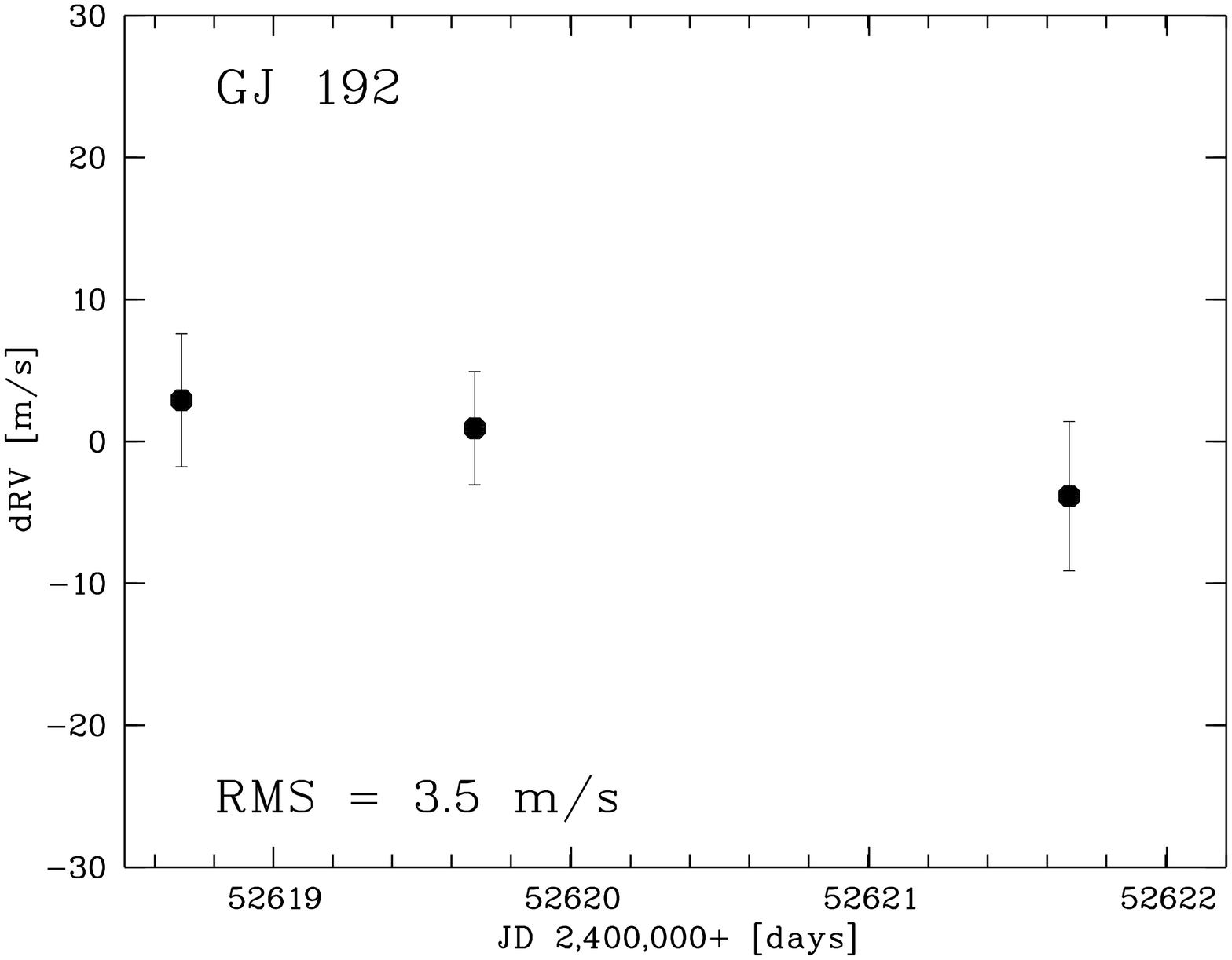,width=9.0cm,height=5.1cm,angle=270}}
        \vbox{\psfig{figure=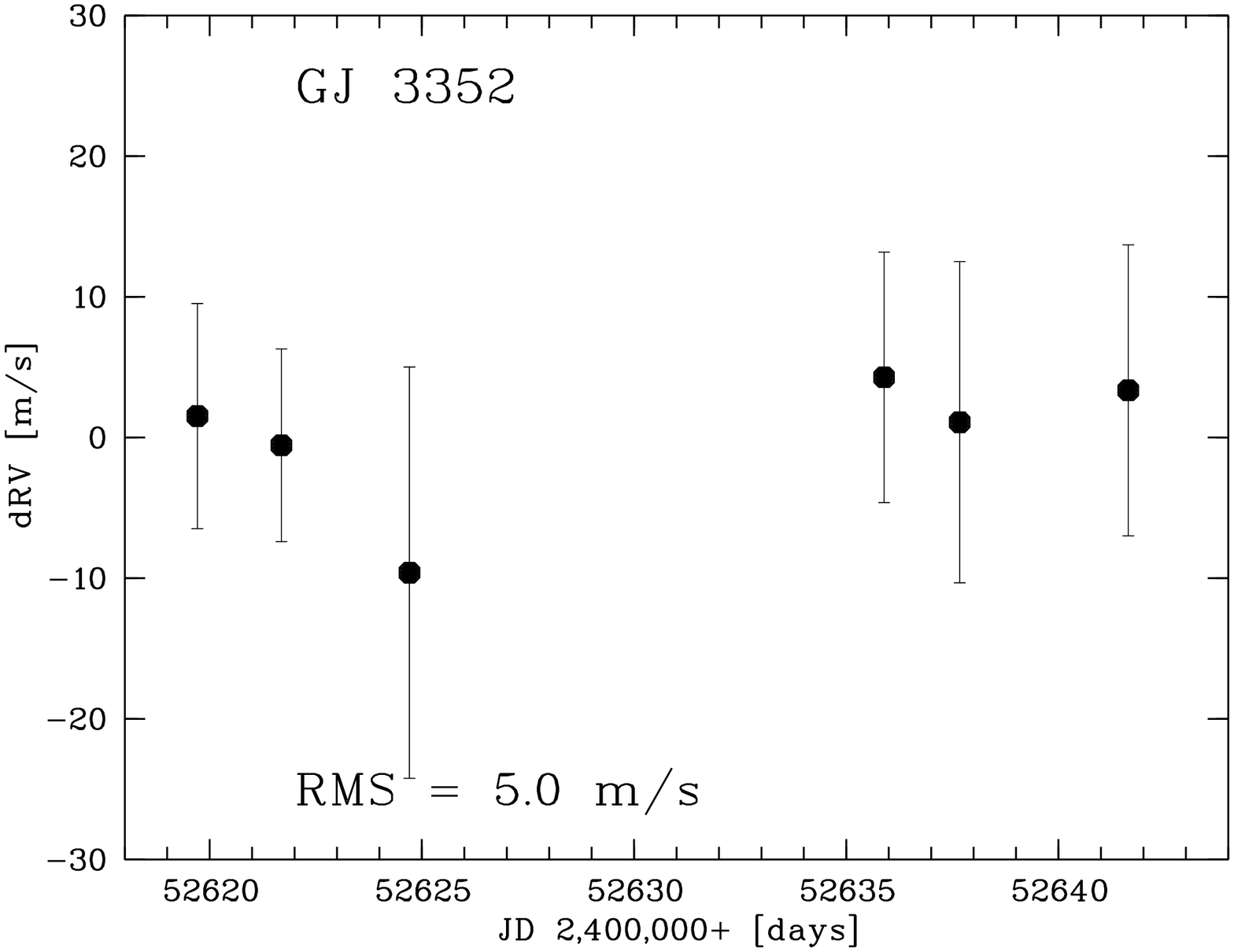,width=9.0cm,height=5.1cm,angle=270}}
        \vbox{\psfig{figure=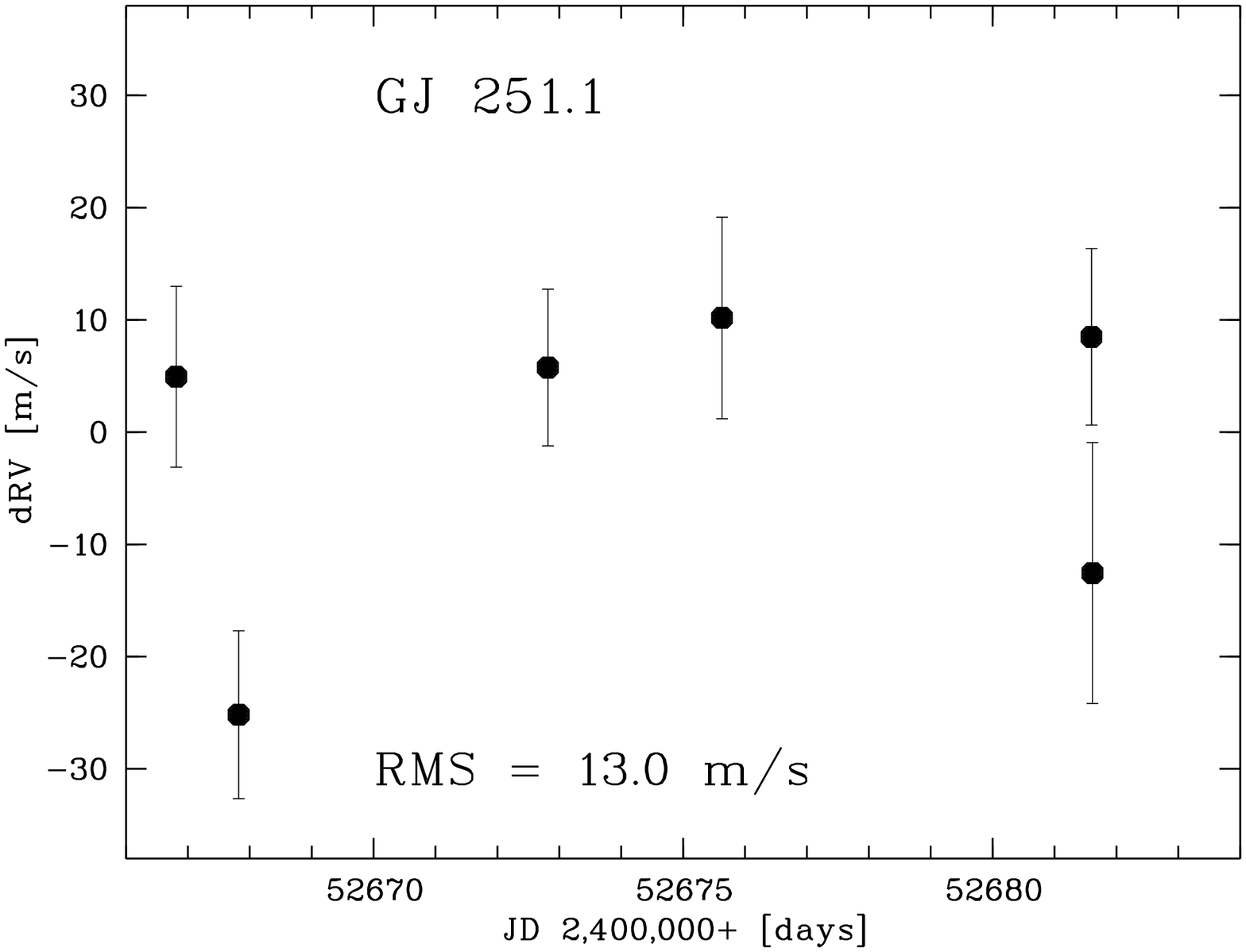,width=9.0cm,height=5.1cm,angle=270}}
        \vbox{\psfig{figure=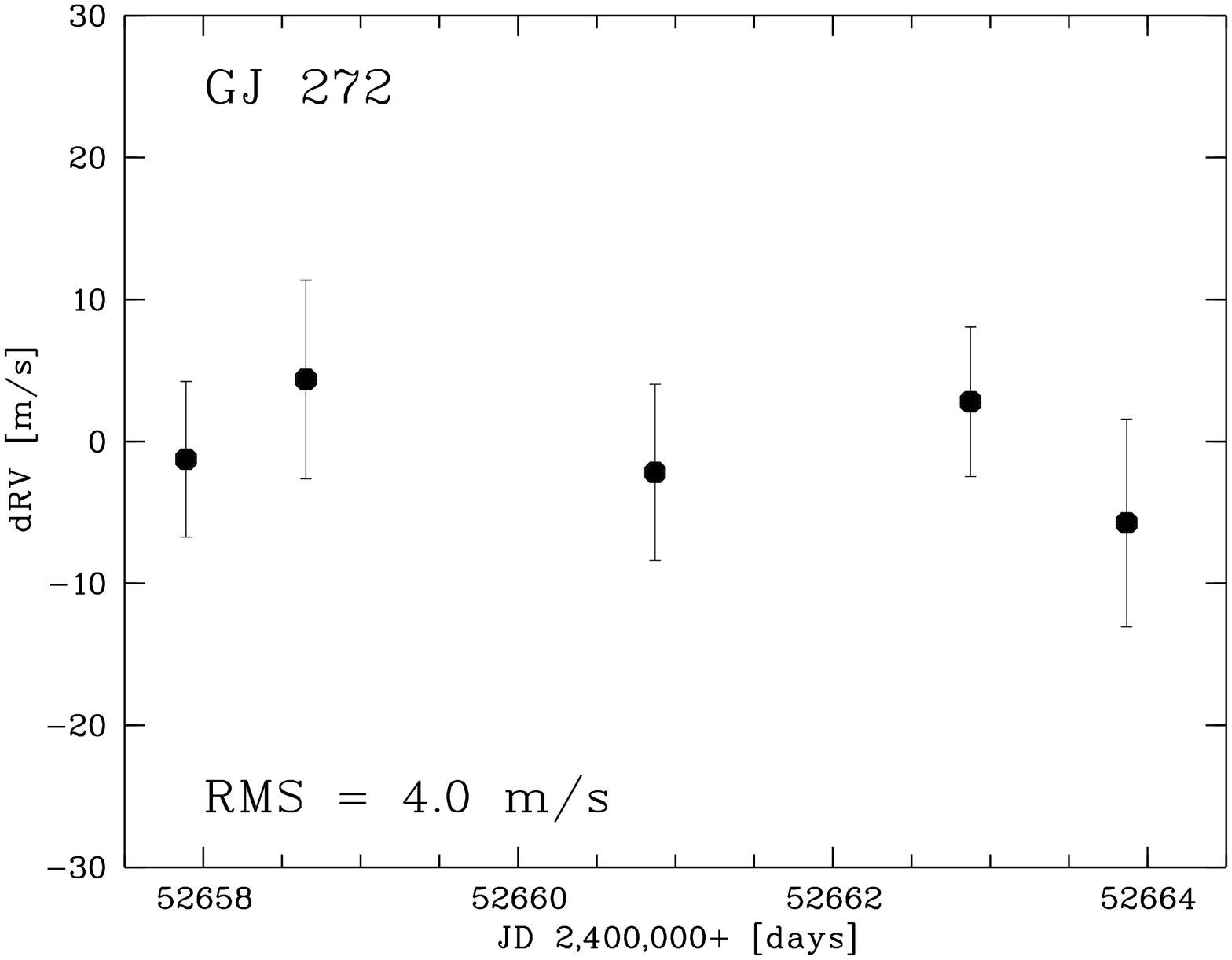,width=9.0cm,height=5.1cm,angle=270}}
   \par
        }
\caption[]{RV results for GJ~192, GJ~3352, GJ~251.1, and GJ~272.}
\label{rvsfig2}
\end{figure}

\begin{figure}
\centering{
        \vbox{\psfig{figure=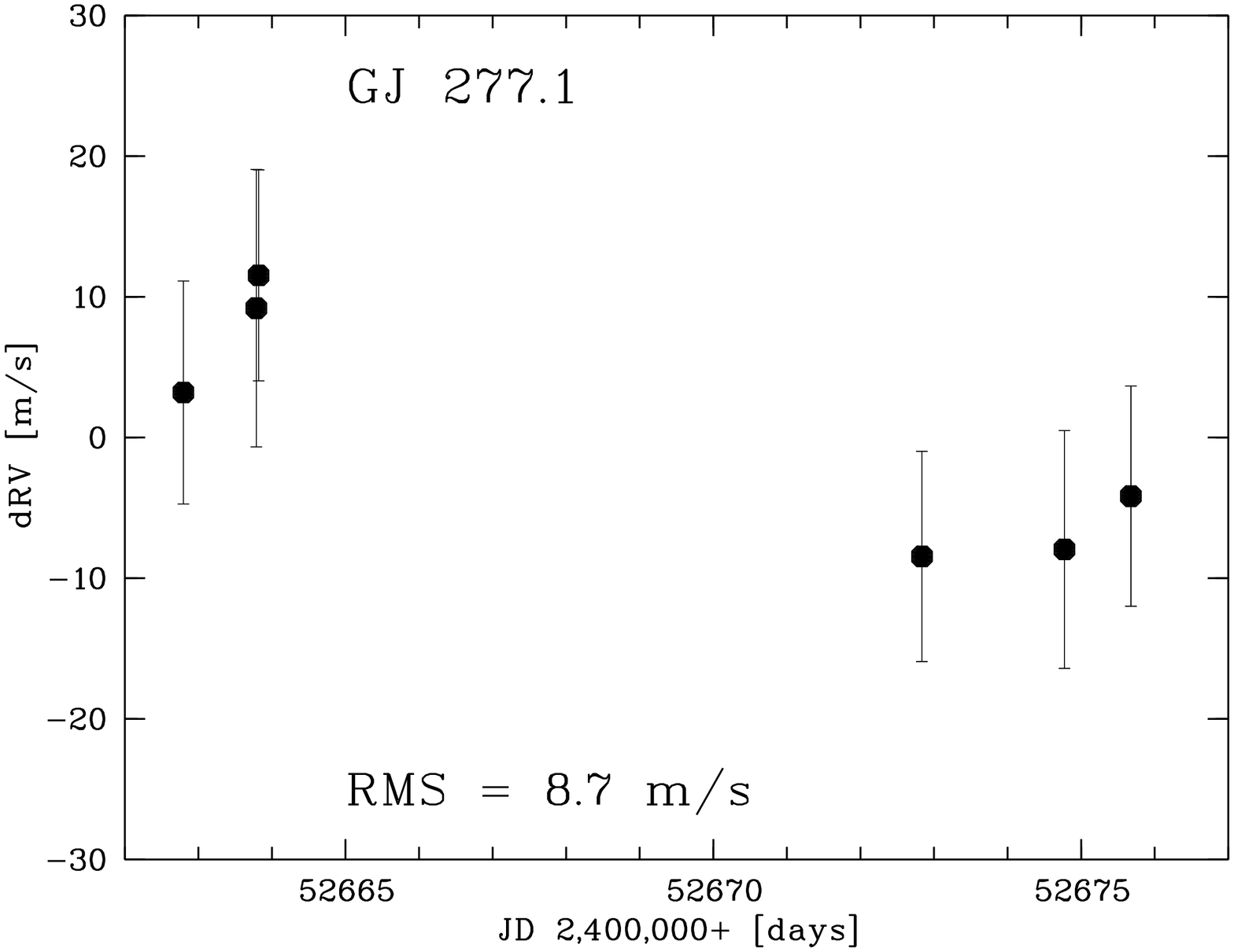,width=9.0cm,height=5.1cm,angle=270}}
        \vbox{\psfig{figure=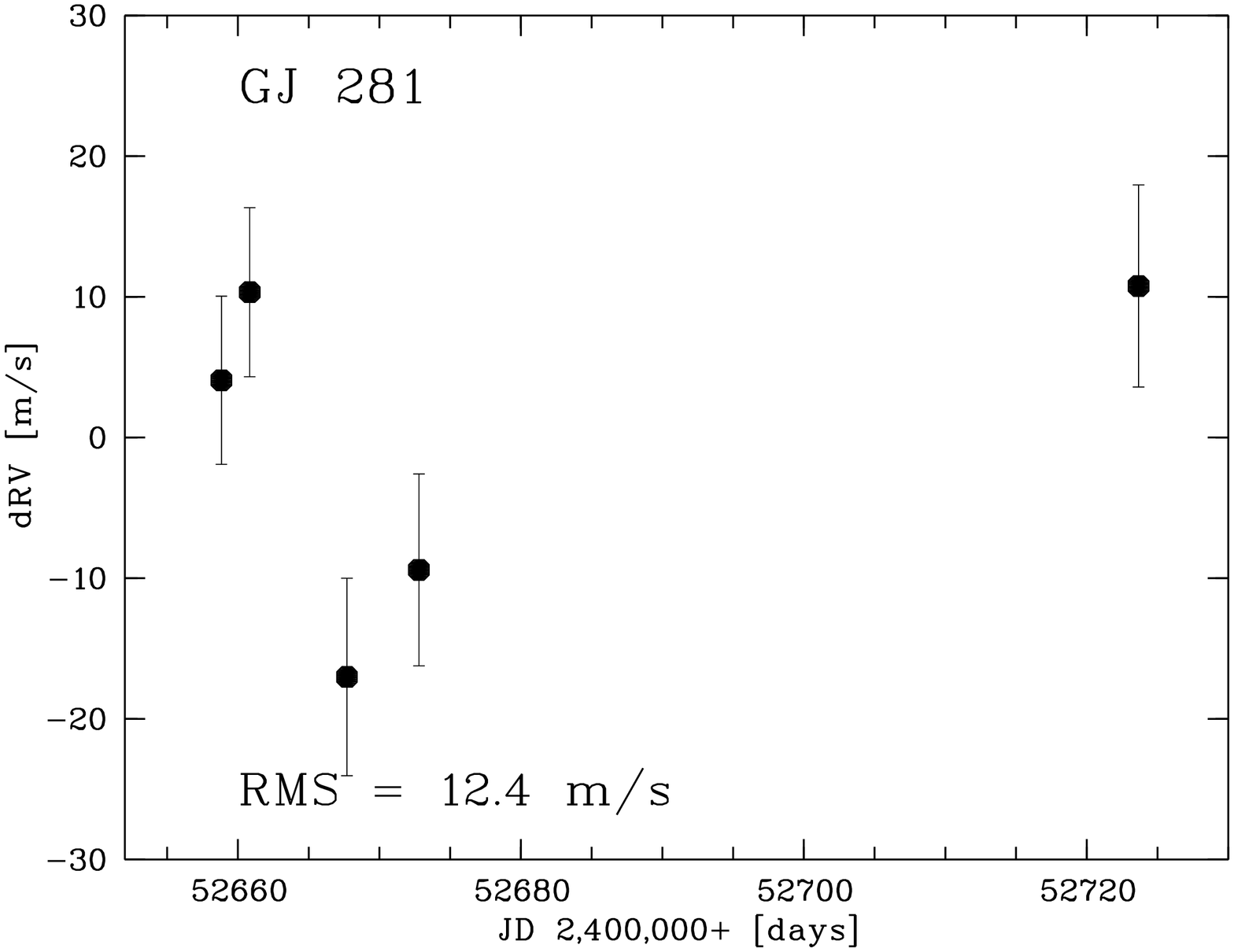,width=9.0cm,height=5.1cm,angle=270}}
        \vbox{\psfig{figure=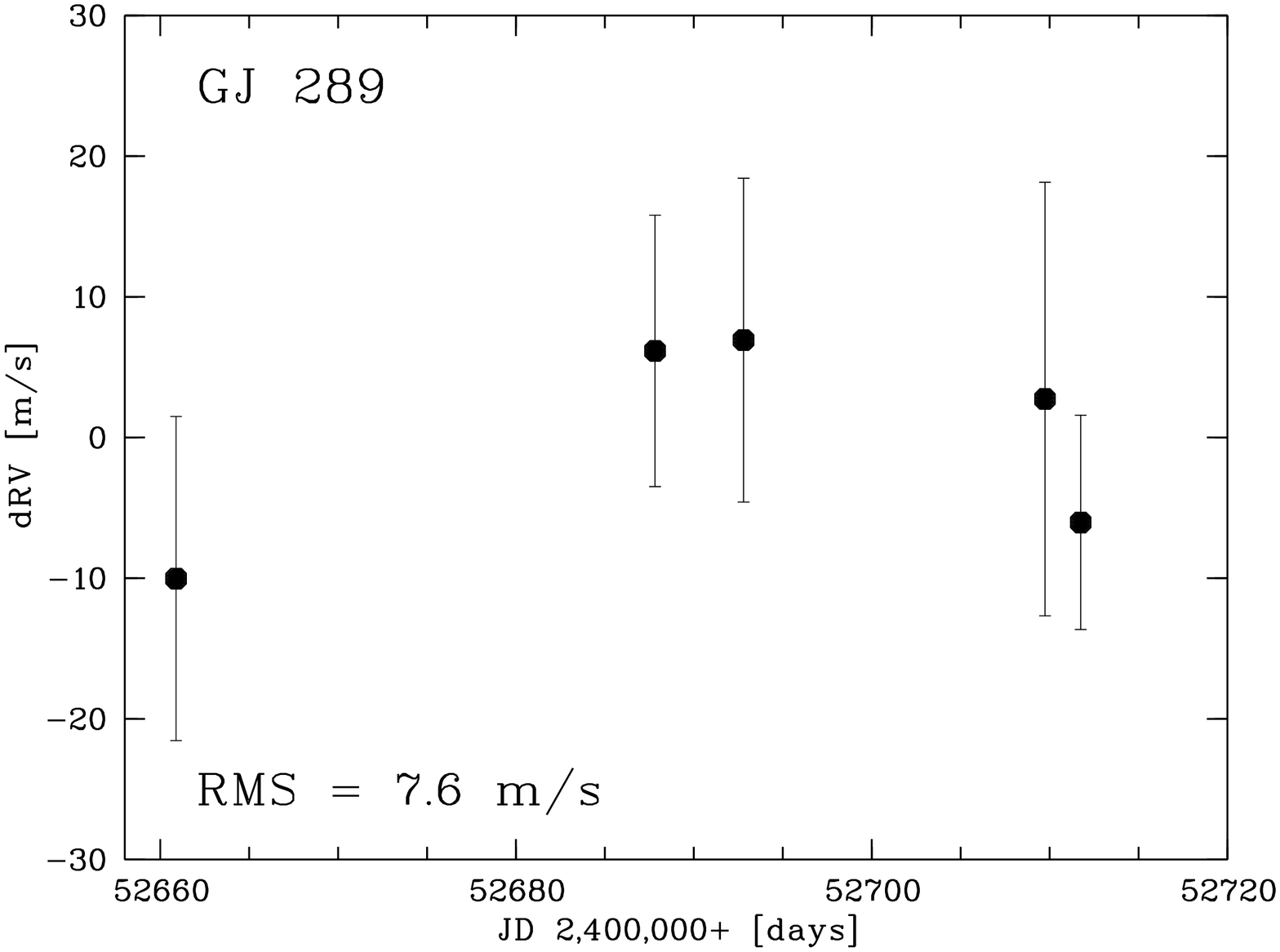,width=9.0cm,height=5.1cm,angle=270}}
        \vbox{\psfig{figure=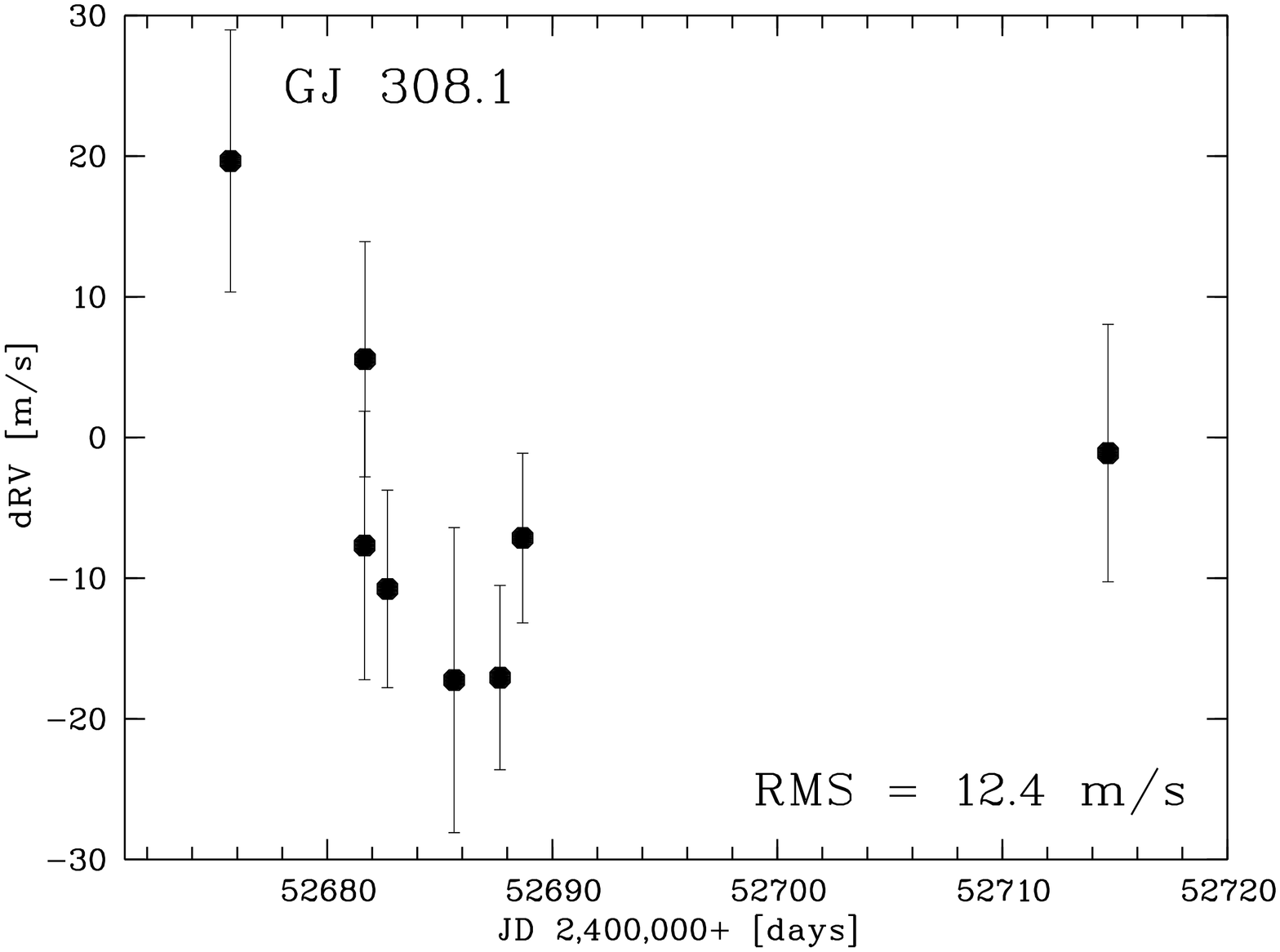,width=9.0cm,height=5.1cm,angle=270}}
   \par
        }
\caption[]{RV results for GJ~277.1, GJ~281, GJ~289, and GJ~308.1.}
\label{rvsfig3}
\end{figure}

\begin{figure}
\centering{
        \vbox{\psfig{figure=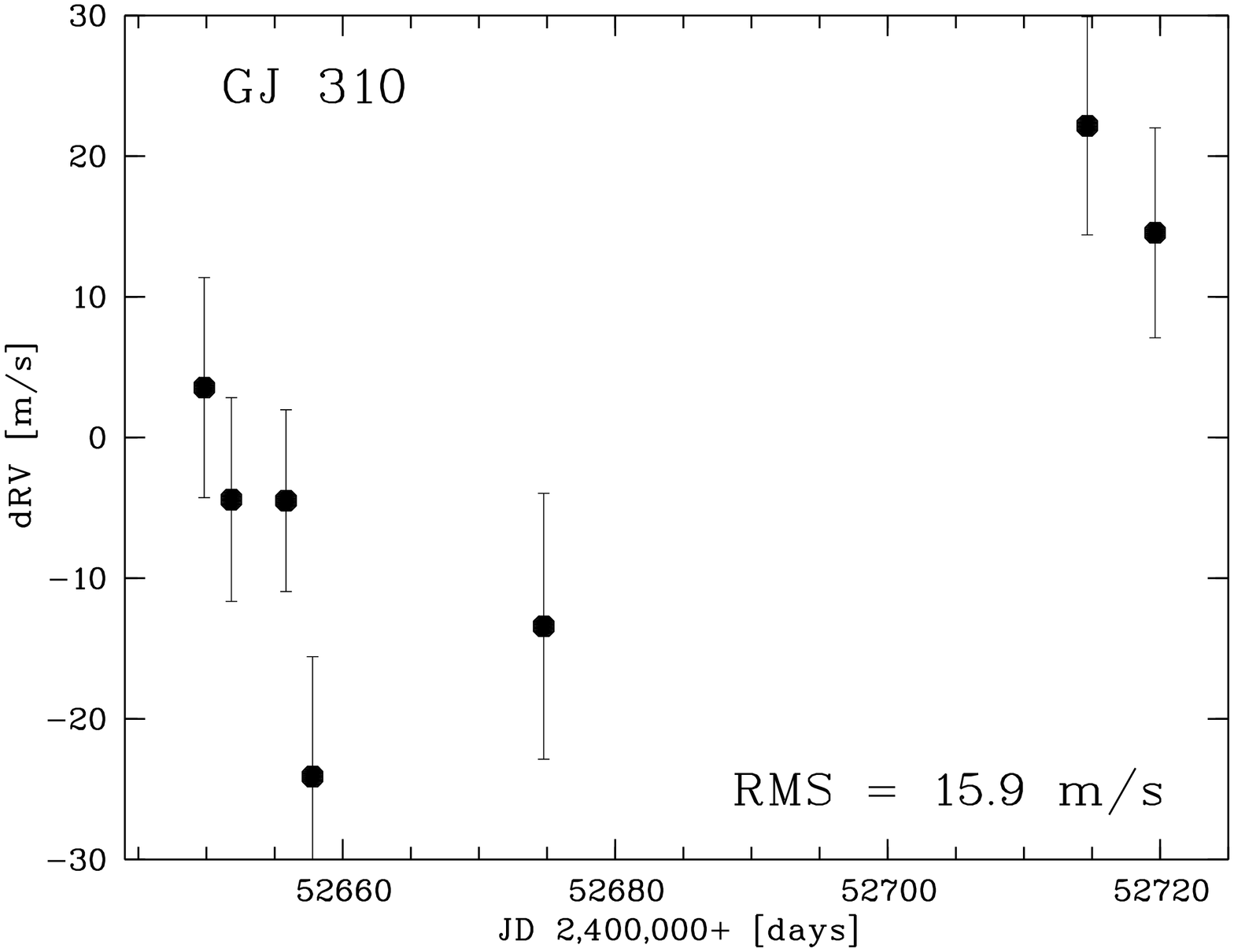,width=9.0cm,height=5.1cm,angle=270}}
        \vbox{\psfig{figure=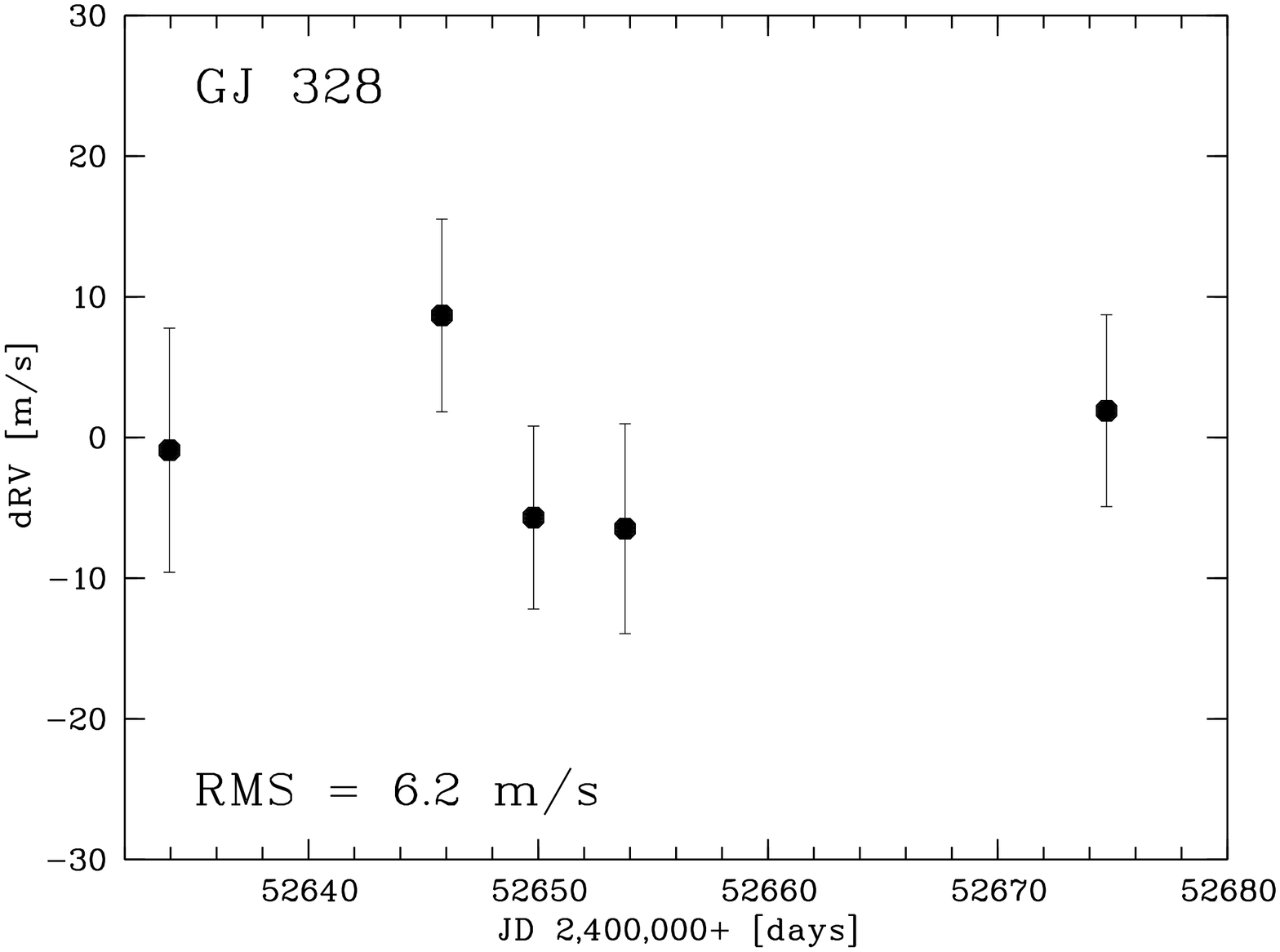,width=9.0cm,height=5.1cm,angle=270}}
        \vbox{\psfig{figure=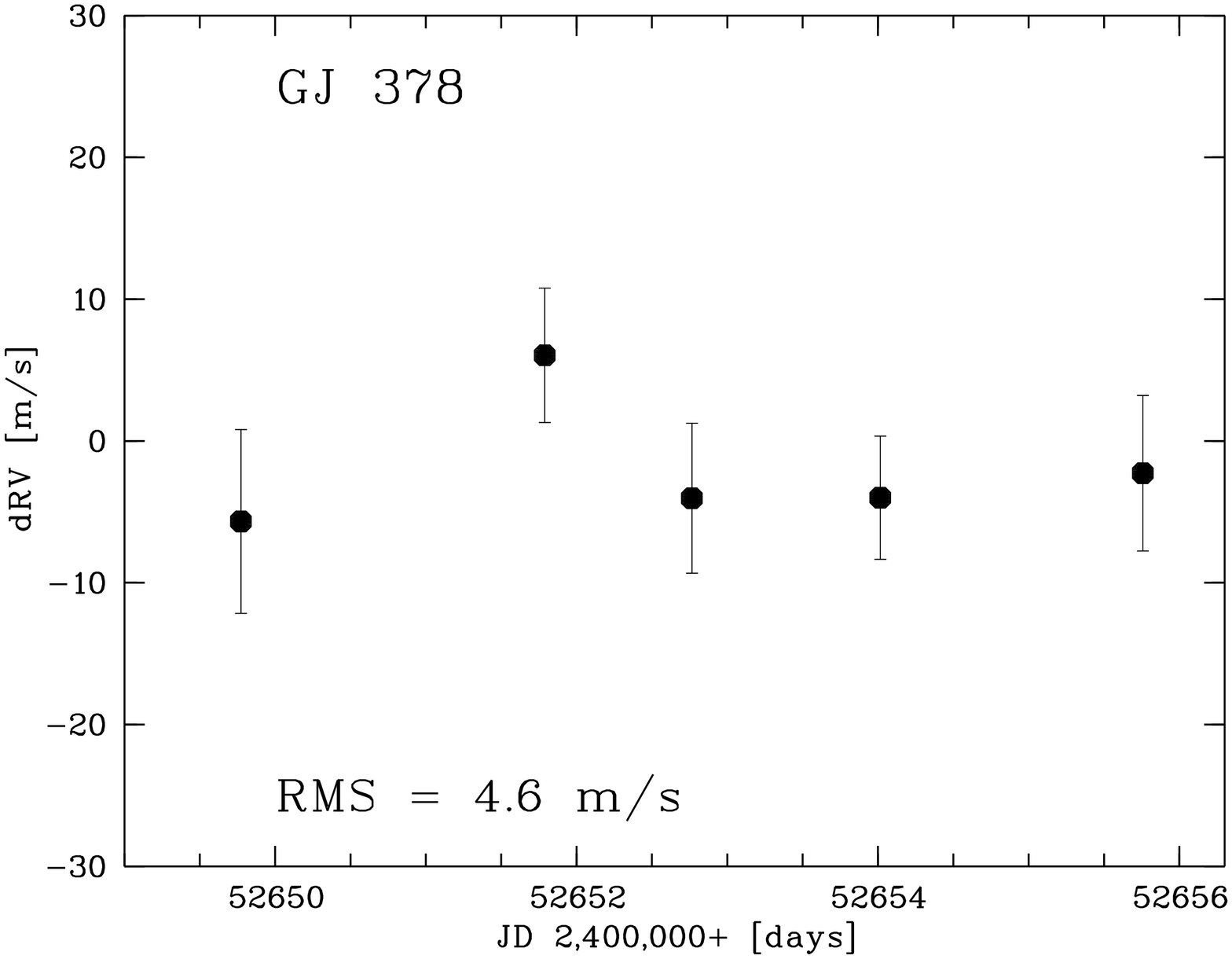,width=9.0cm,height=5.1cm,angle=270}}
        \vbox{\psfig{figure=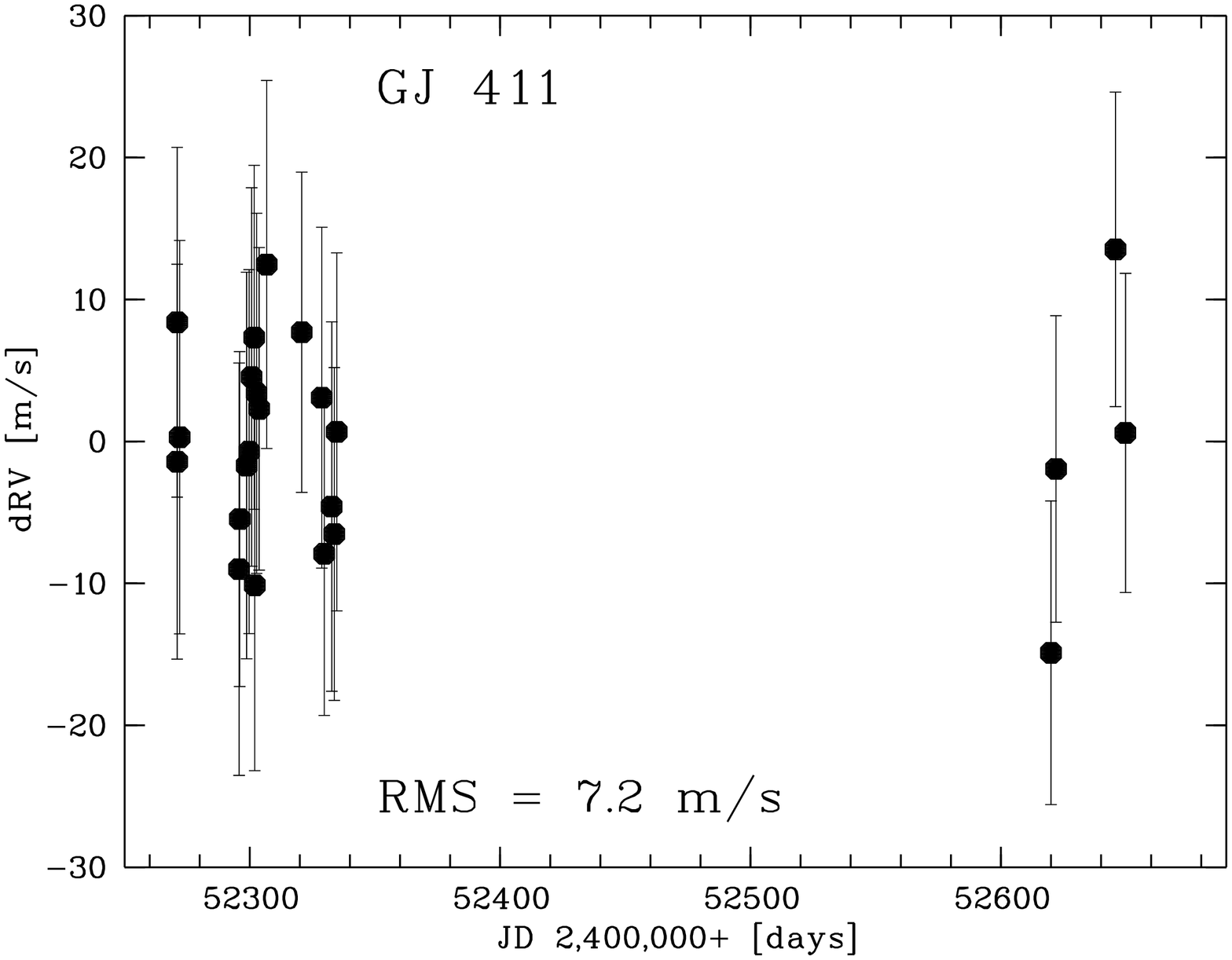,width=9.0cm,height=5.1cm,angle=270}}
   \par
        }
\caption[]{RV results for GJ~310, GJ~328, GJ~353, GJ~378, and GJ~411}
\label{rvsfig4}
\end{figure}

\begin{figure}
\centering{
        \vbox{\psfig{figure=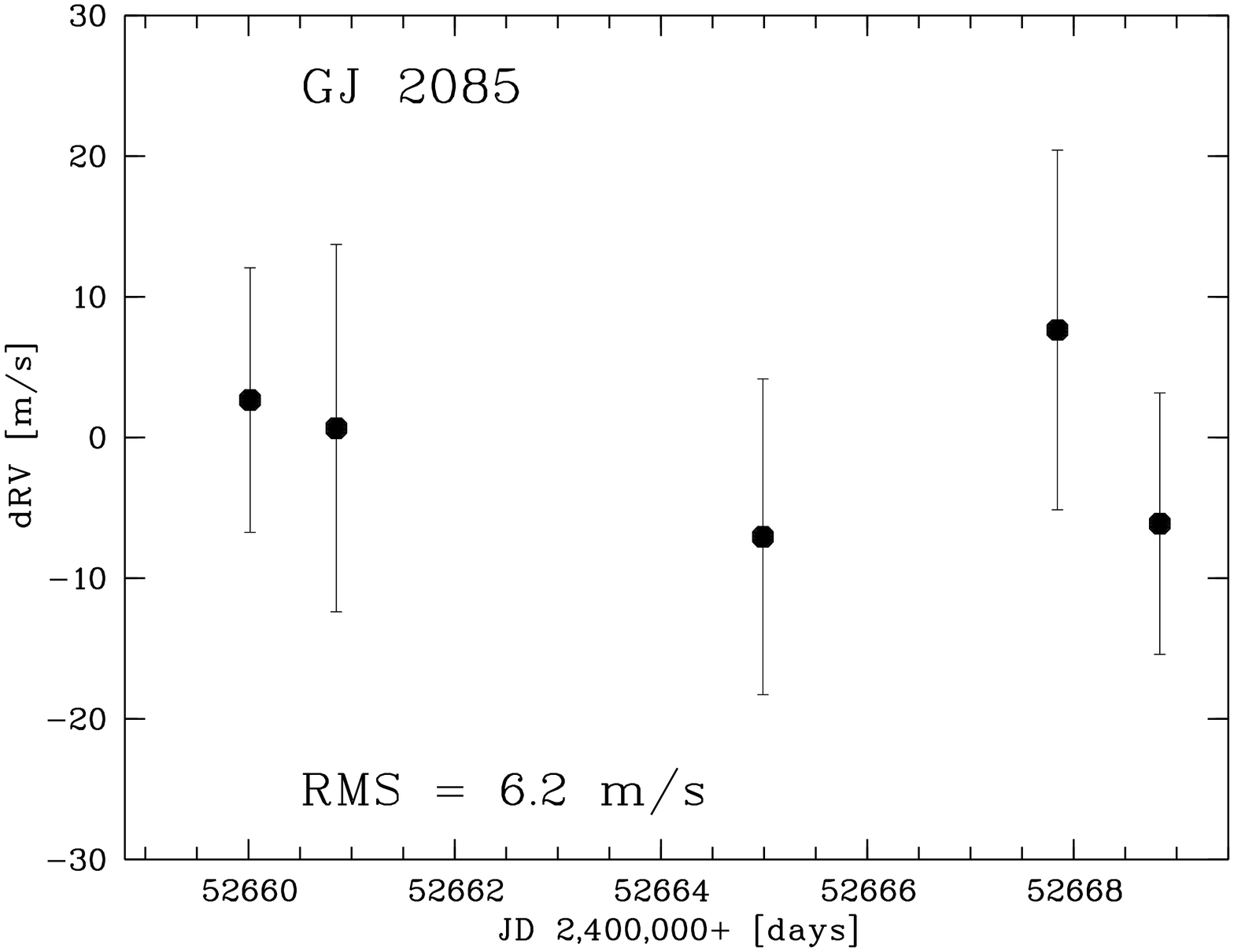,width=9.0cm,height=5.1cm,angle=270}}
        \vbox{\psfig{figure=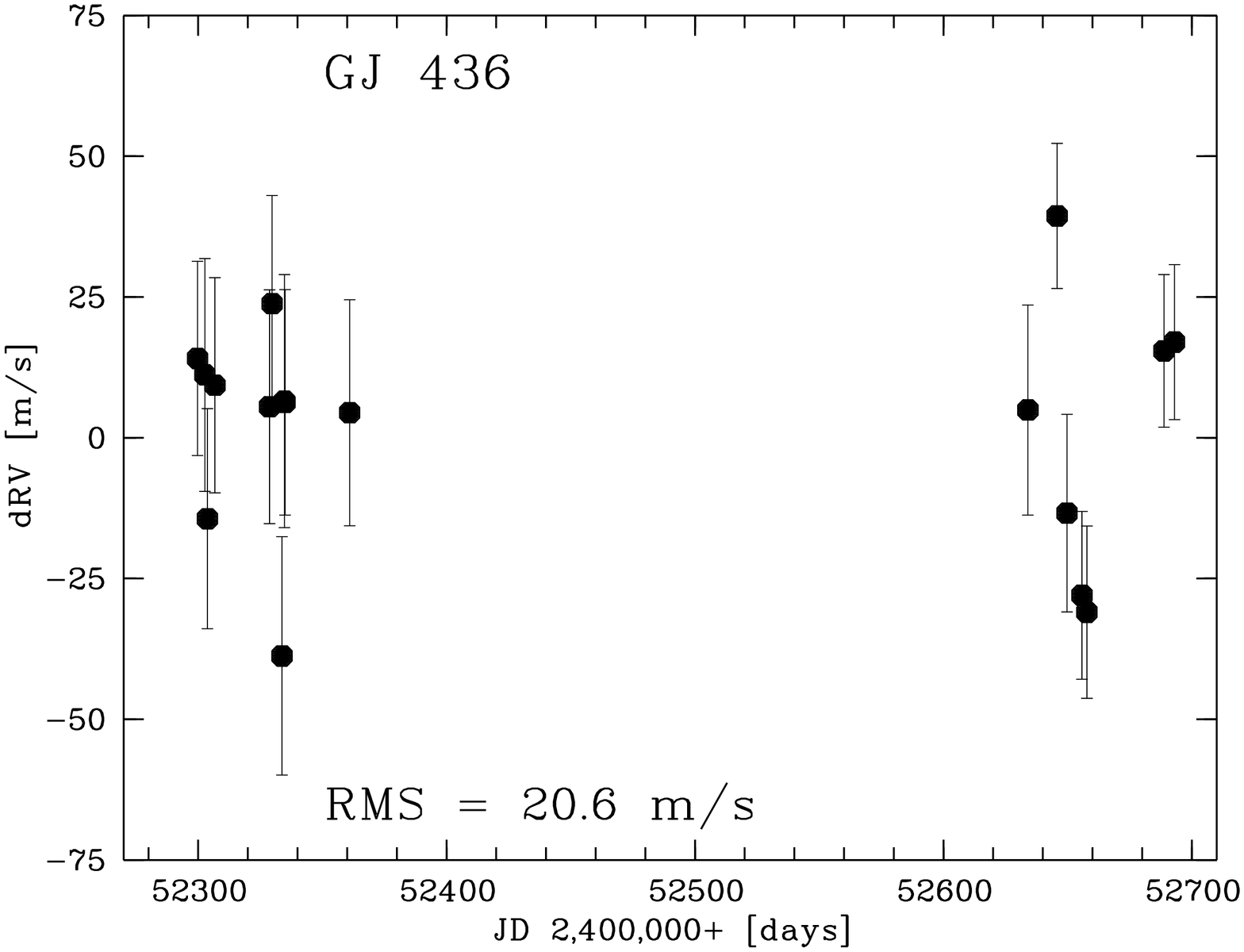,width=9.0cm,height=5.1cm,angle=270}}
        \vbox{\psfig{figure=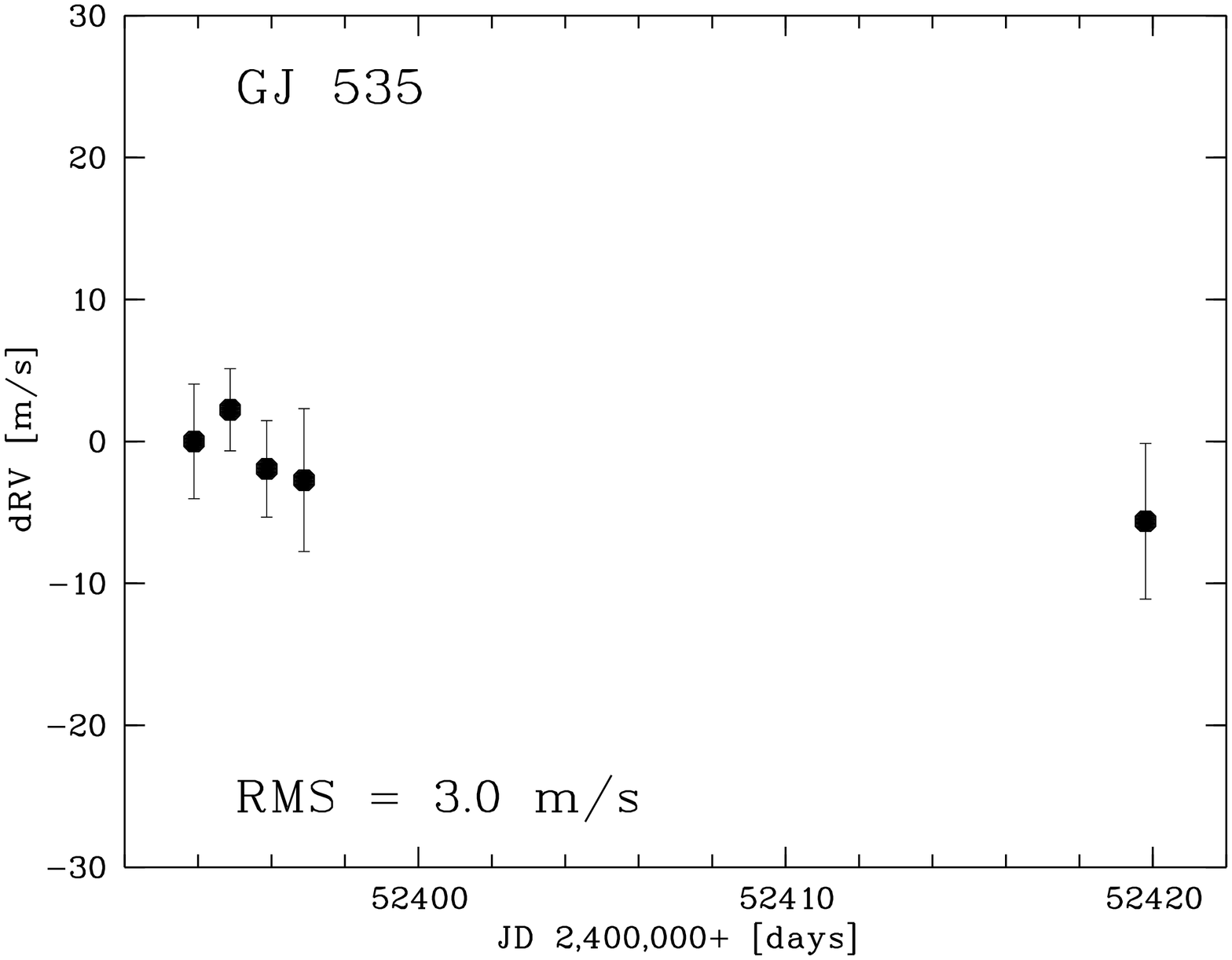,width=9.0cm,height=5.1cm,angle=270}}
        \vbox{\psfig{figure=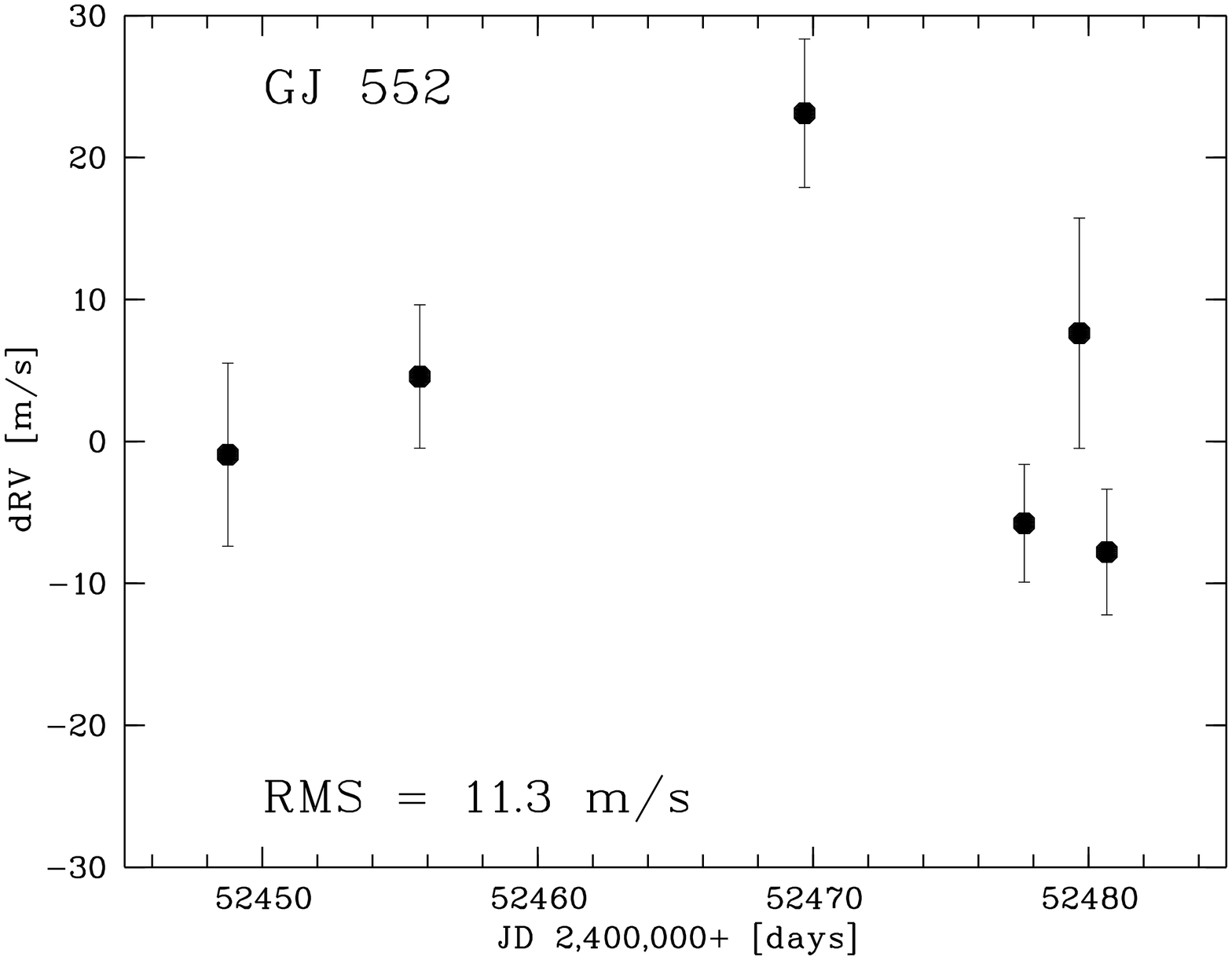,width=9.0cm,height=5.1cm,angle=270}}
   \par
        }
\caption[]{RV results for GJ~2085, GJ~436 (note the larger velocity scale), GJ~535, and GJ~552.}
\label{rvsfig5}
\end{figure}

\begin{figure}
\centering{
        \vbox{\psfig{figure=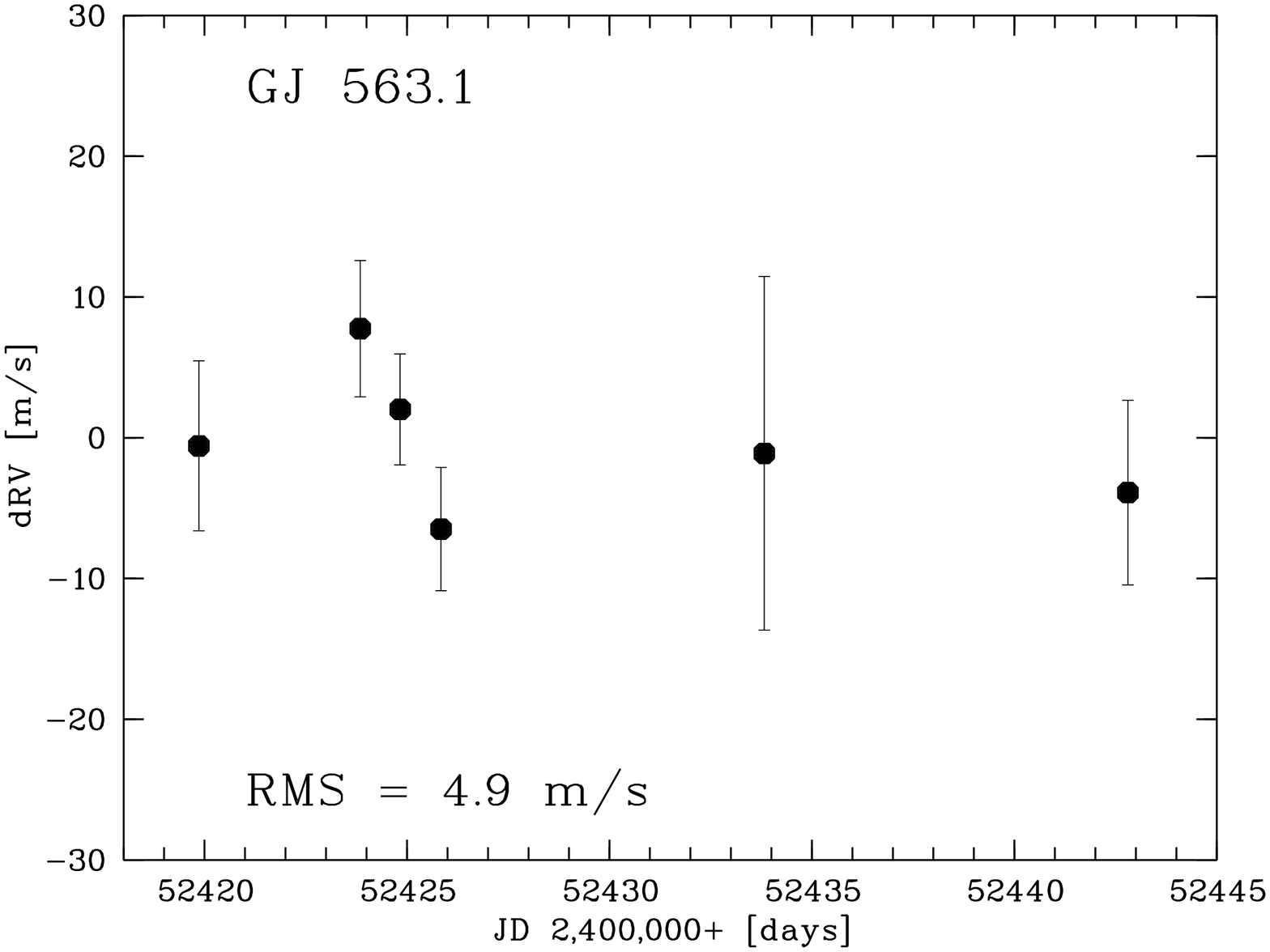,width=9.0cm,height=5.1cm,angle=270}}
        \vbox{\psfig{figure=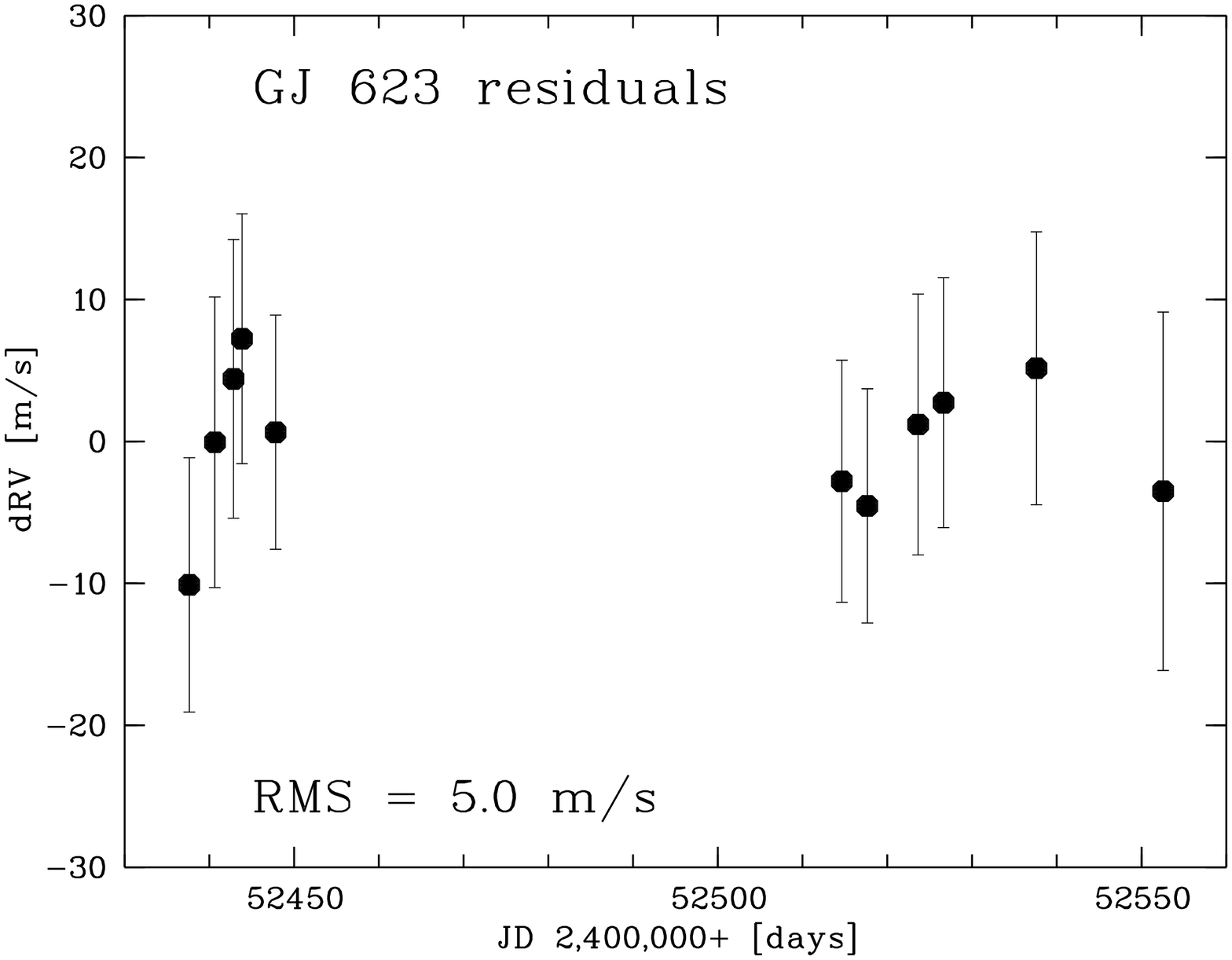,width=9.0cm,height=5.1cm,angle=270}}
        \vbox{\psfig{figure=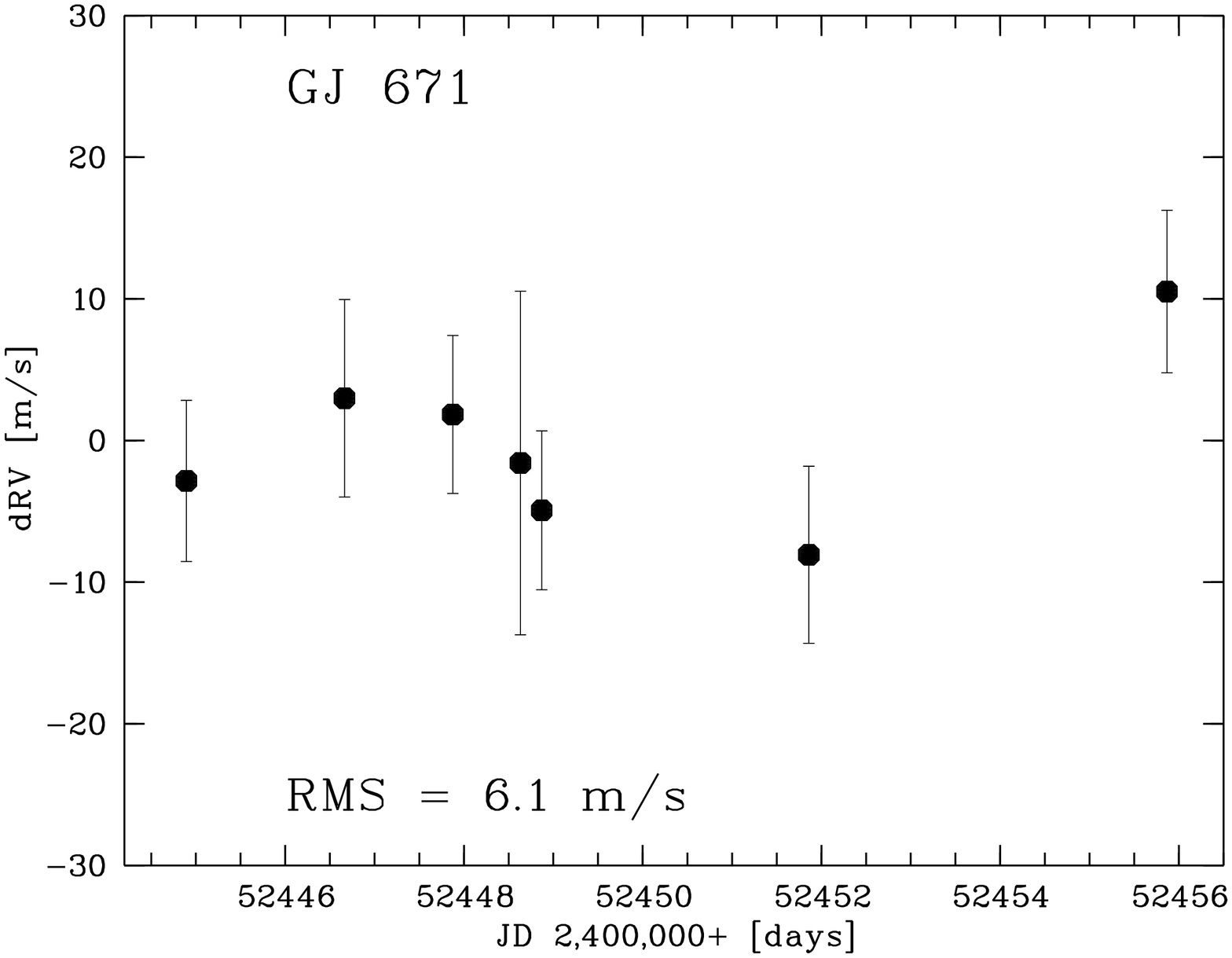,width=9.0cm,height=5.1cm,angle=270}}
        \vbox{\psfig{figure=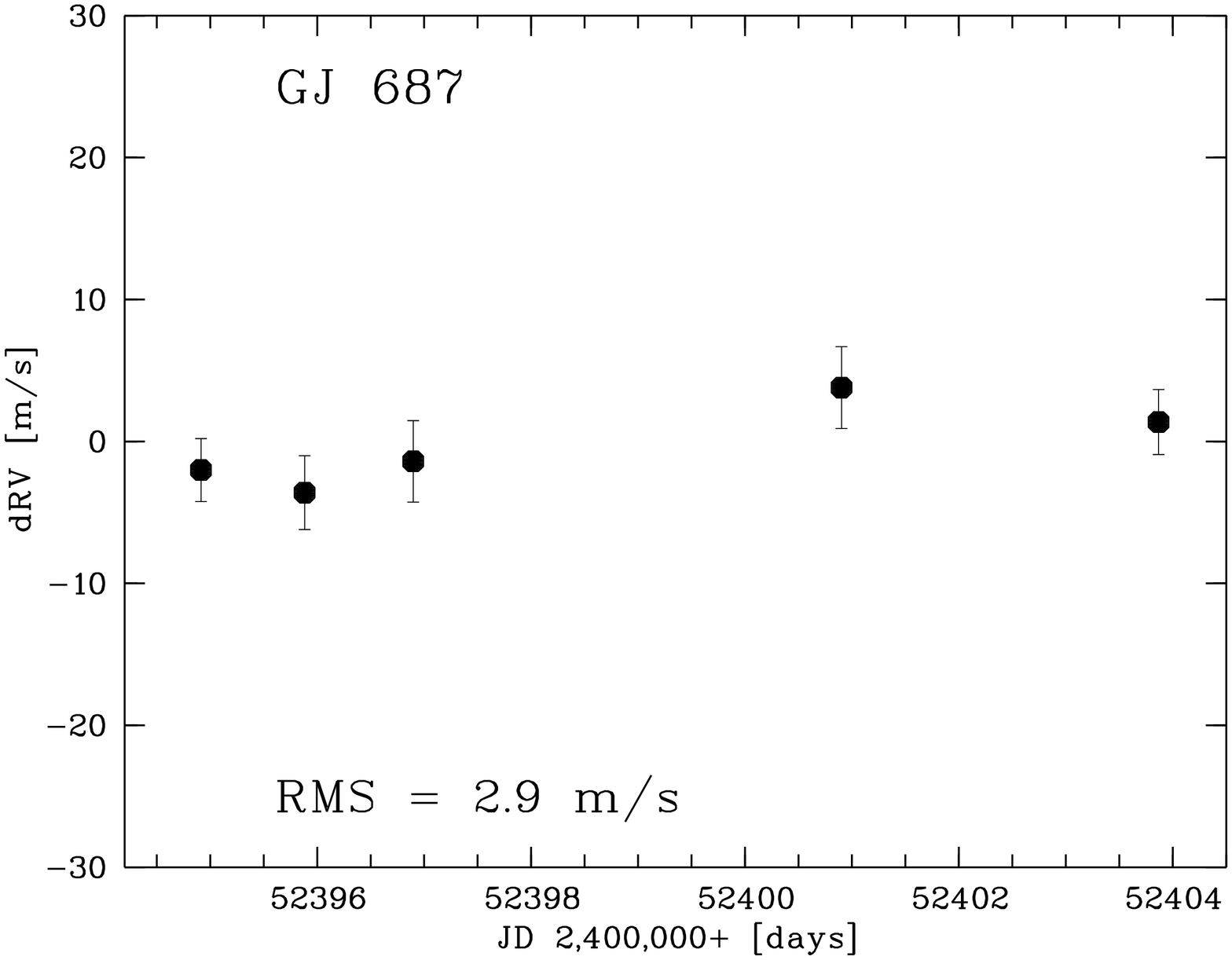,width=9.0cm,height=5.1cm,angle=270}}
   \par
        }
\caption[]{RV results for GJ~563.1, GJ~623 (residuals), GJ~671, and GJ~687}
\label{rvsfig6}
\end{figure}

\begin{figure}
\centering{
        \vbox{\psfig{figure=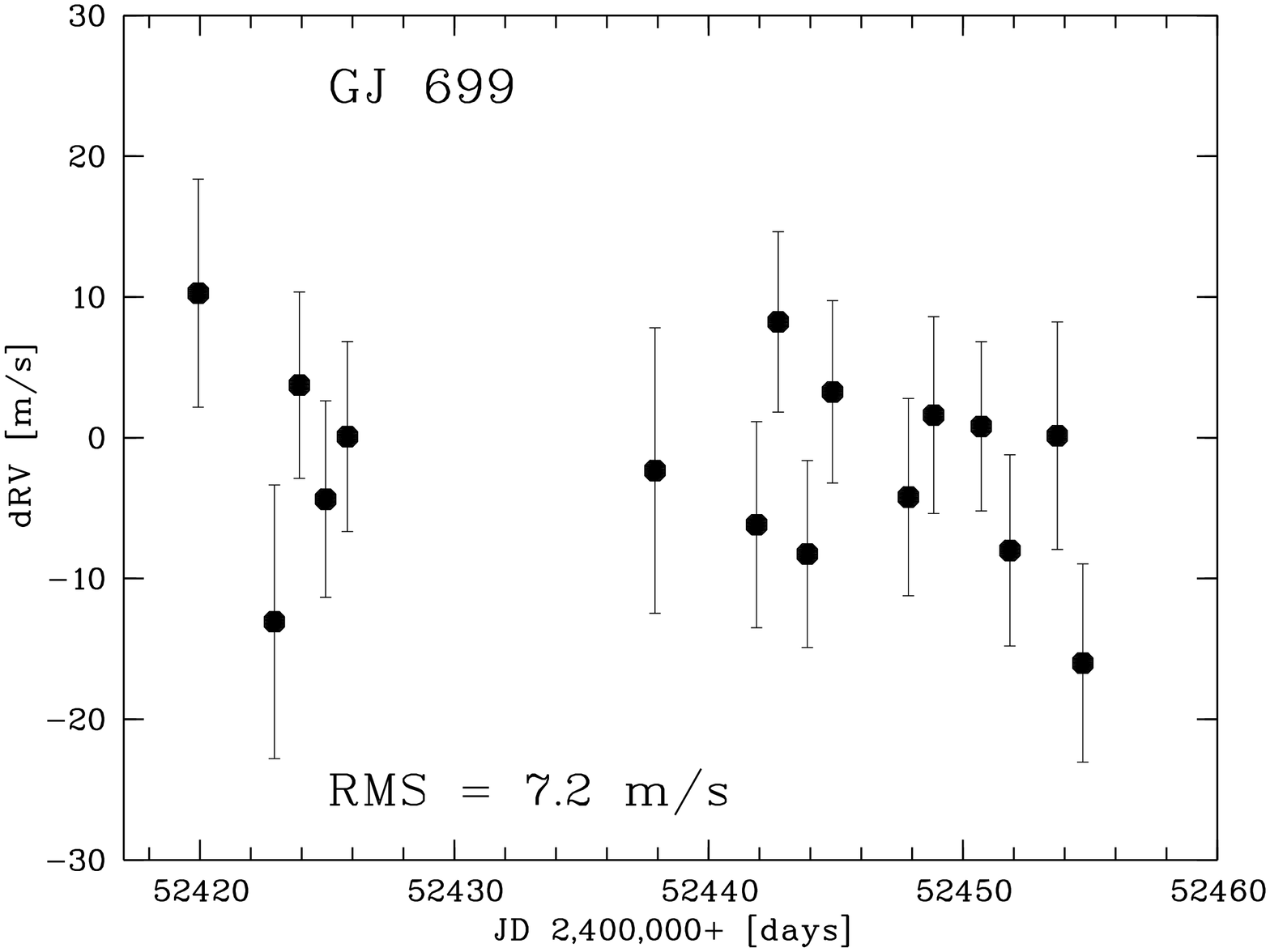,width=9.0cm,height=5.1cm,angle=270}}
        \vbox{\psfig{figure=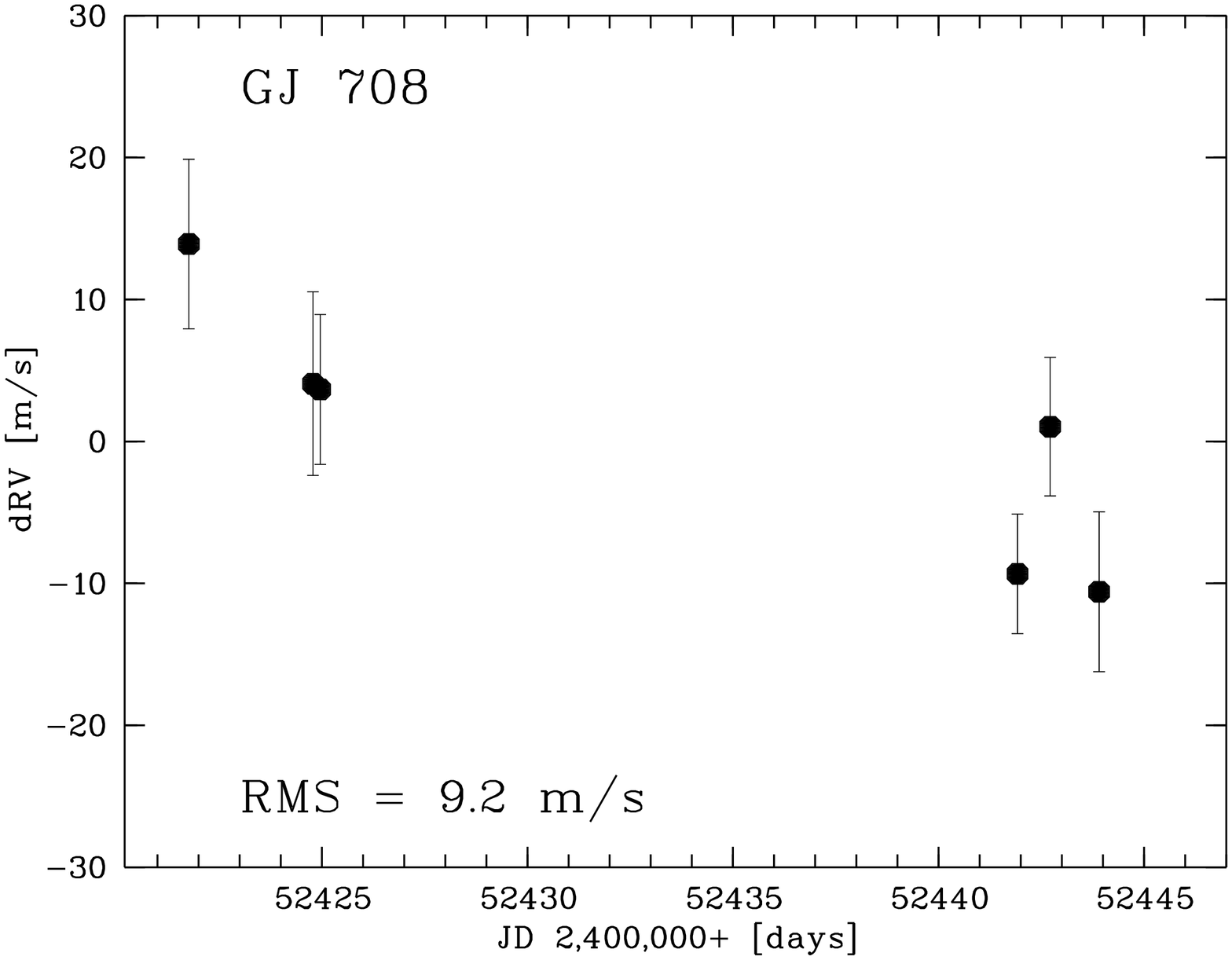,width=9.0cm,height=5.1cm,angle=270}}
        \vbox{\psfig{figure=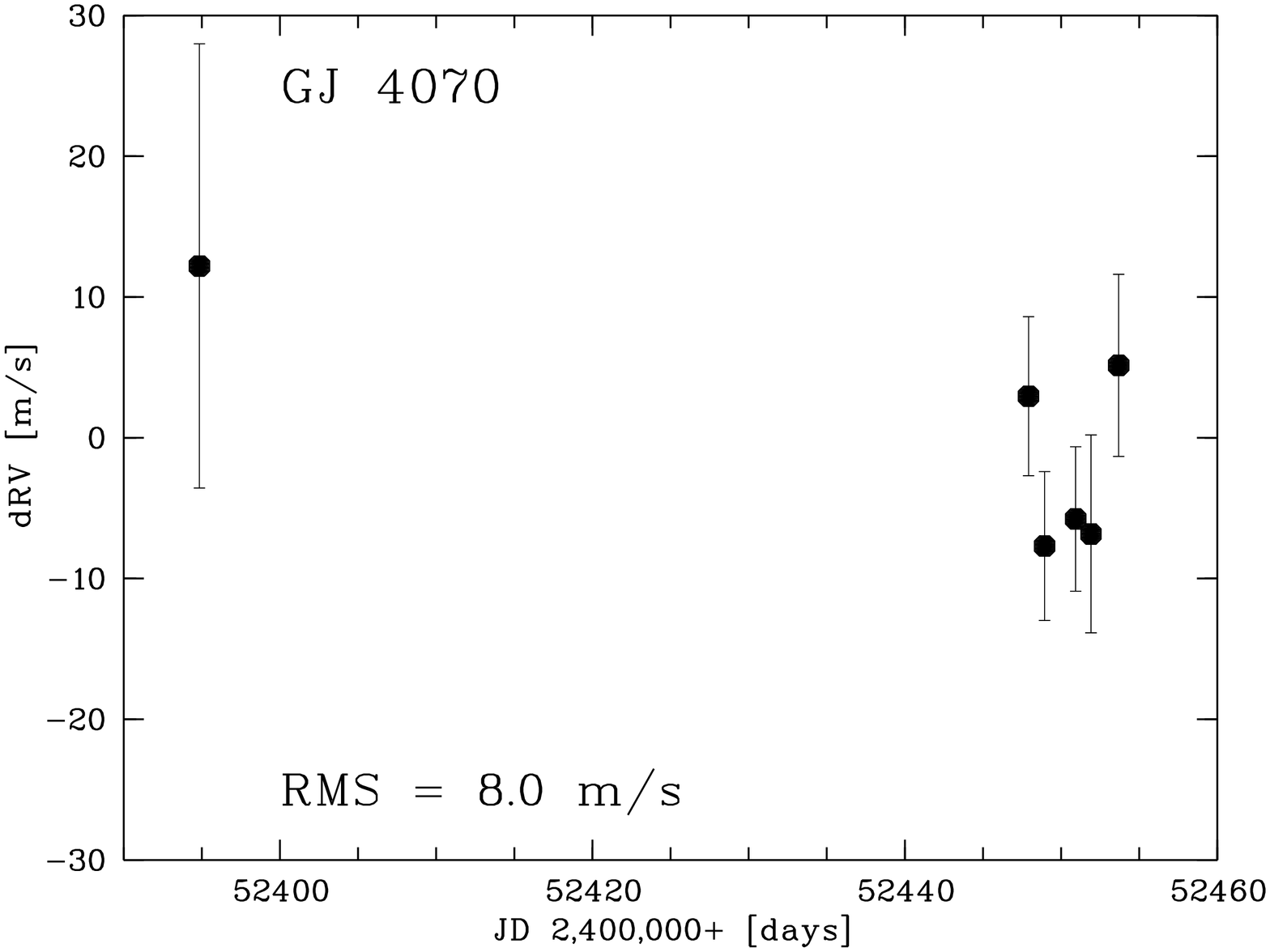,width=9.0cm,height=5.1cm,angle=270}}
        \vbox{\psfig{figure=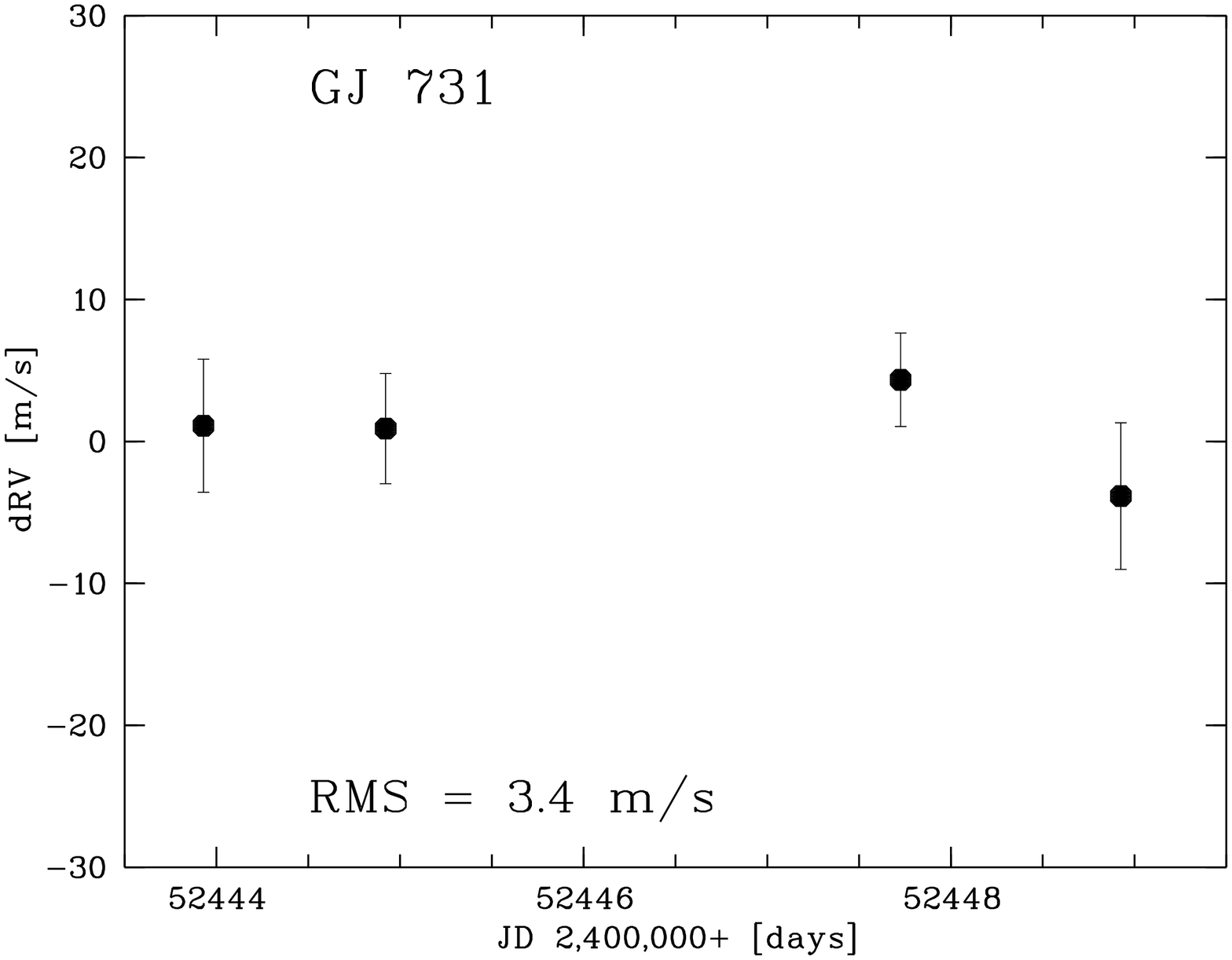,width=9.0cm,height=5.1cm,angle=270}}
   \par
        }
\caption[]{RV results for GJ~699, GJ~708, GJ~4070, and GJ~731.}
\label{rvsfig7}
\end{figure}

\begin{figure}
\centering{
        \vbox{\psfig{figure=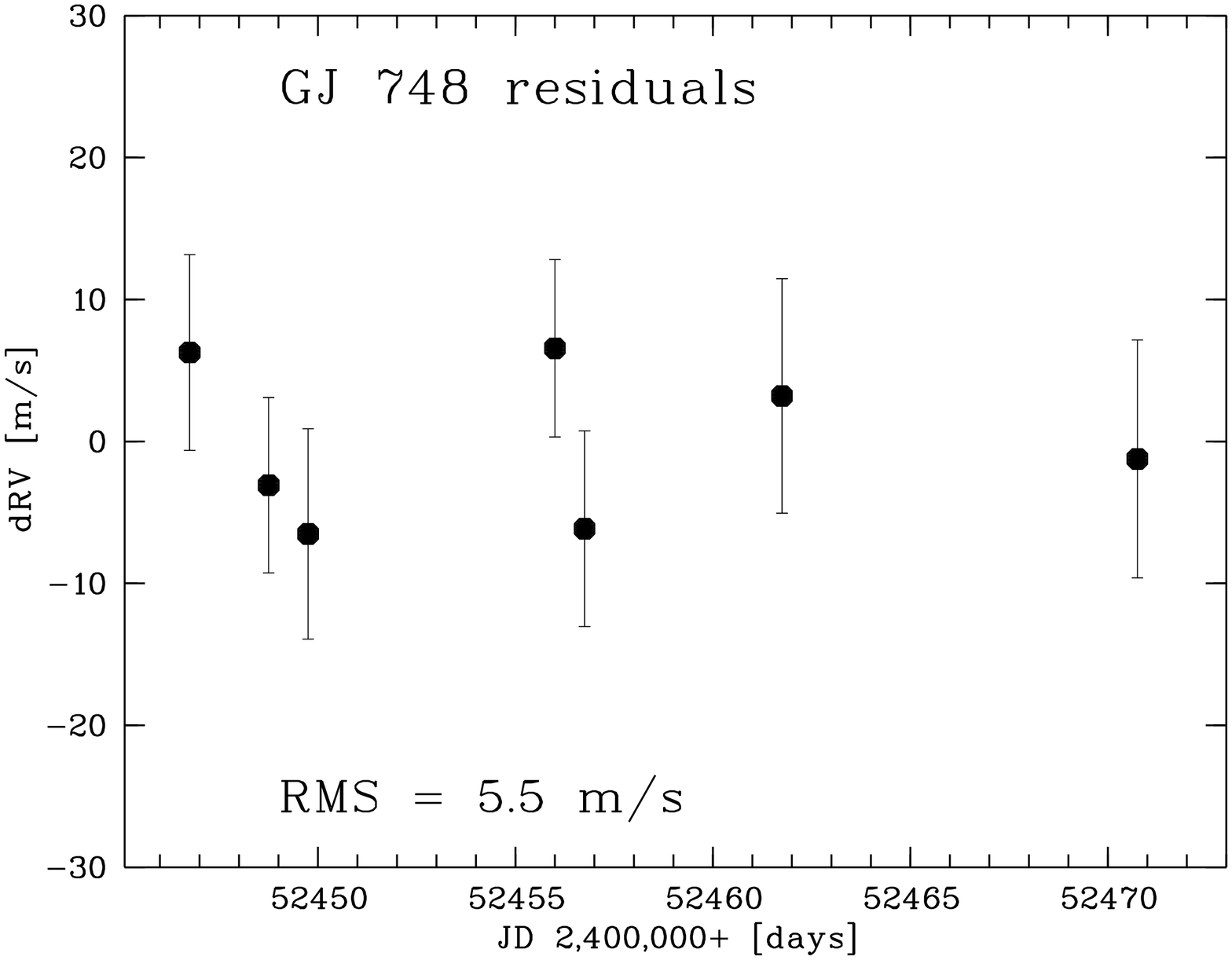,width=9.0cm,height=5.1cm,angle=270}}
        \vbox{\psfig{figure=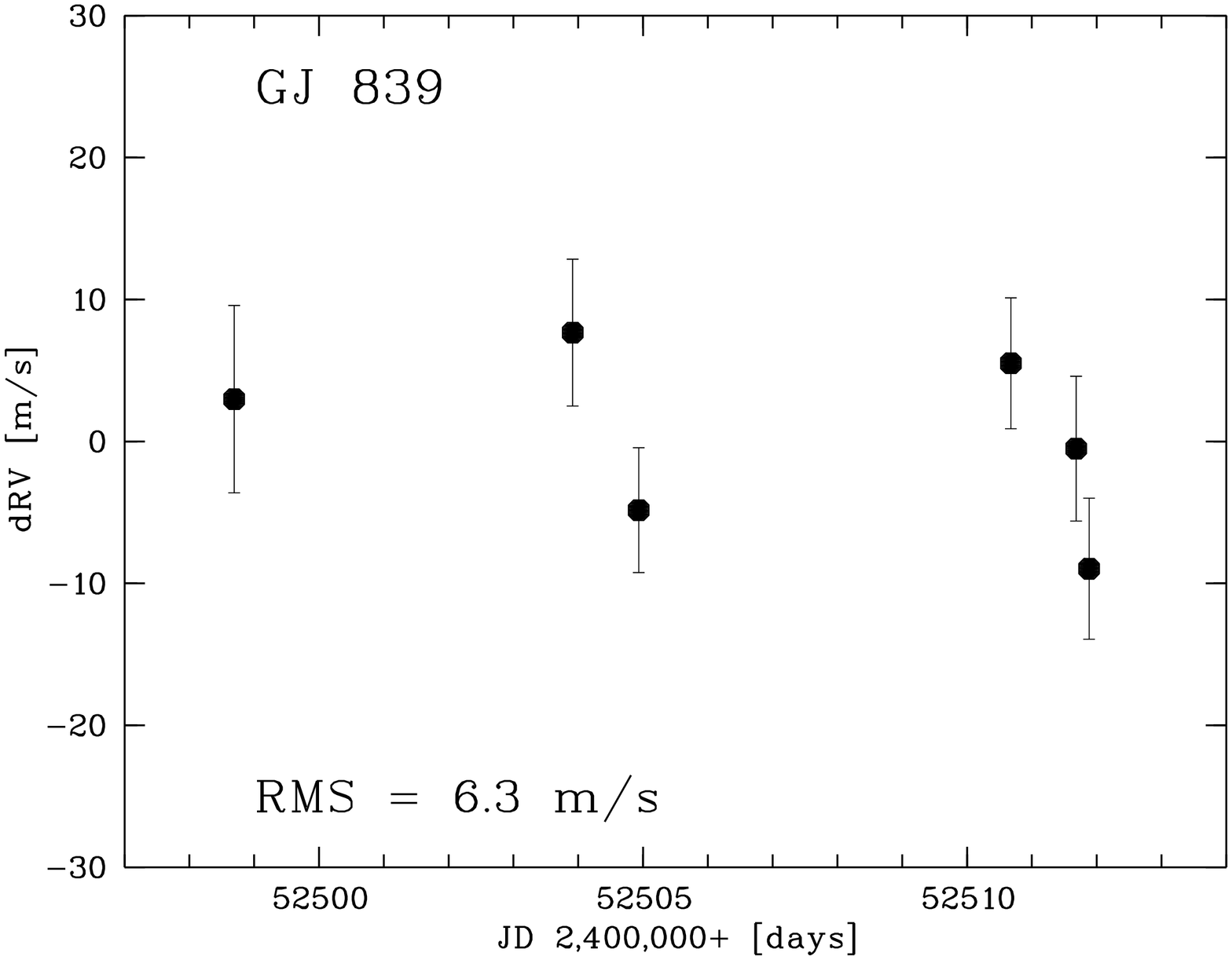,width=9.0cm,height=5.1cm,angle=270}}
        \vbox{\psfig{figure=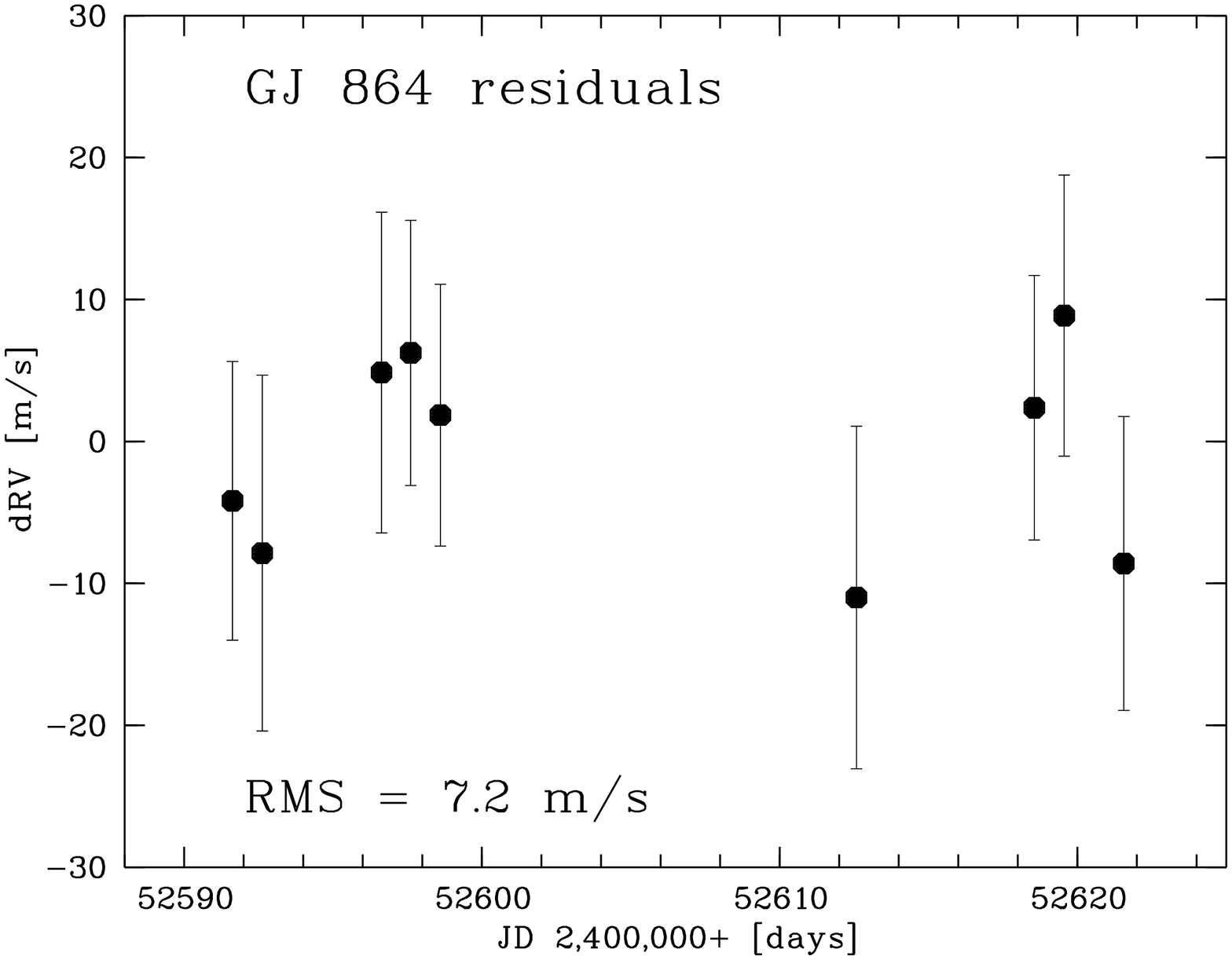,width=9.0cm,height=5.1cm,angle=270}}
        \vbox{\psfig{figure=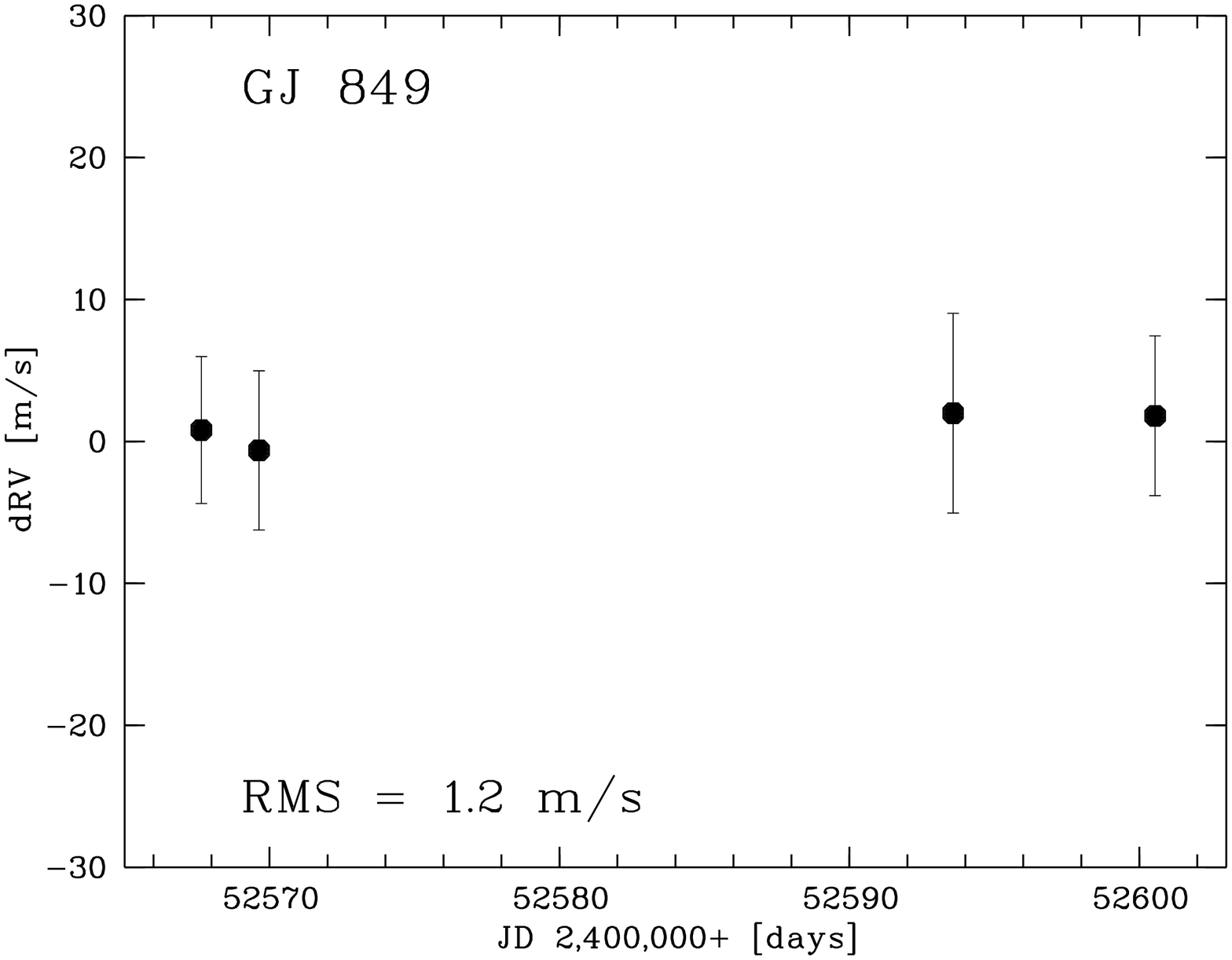,width=9.0cm,height=5.1cm,angle=270}}
   \par
        }
\caption[]{RV results for GJ~748 (residuals), GJ~839, GJ~864 (residuals), and GJ~849.}
\label{rvsfig8}
\end{figure}

\begin{figure}
\centering{
        \vbox{\psfig{figure=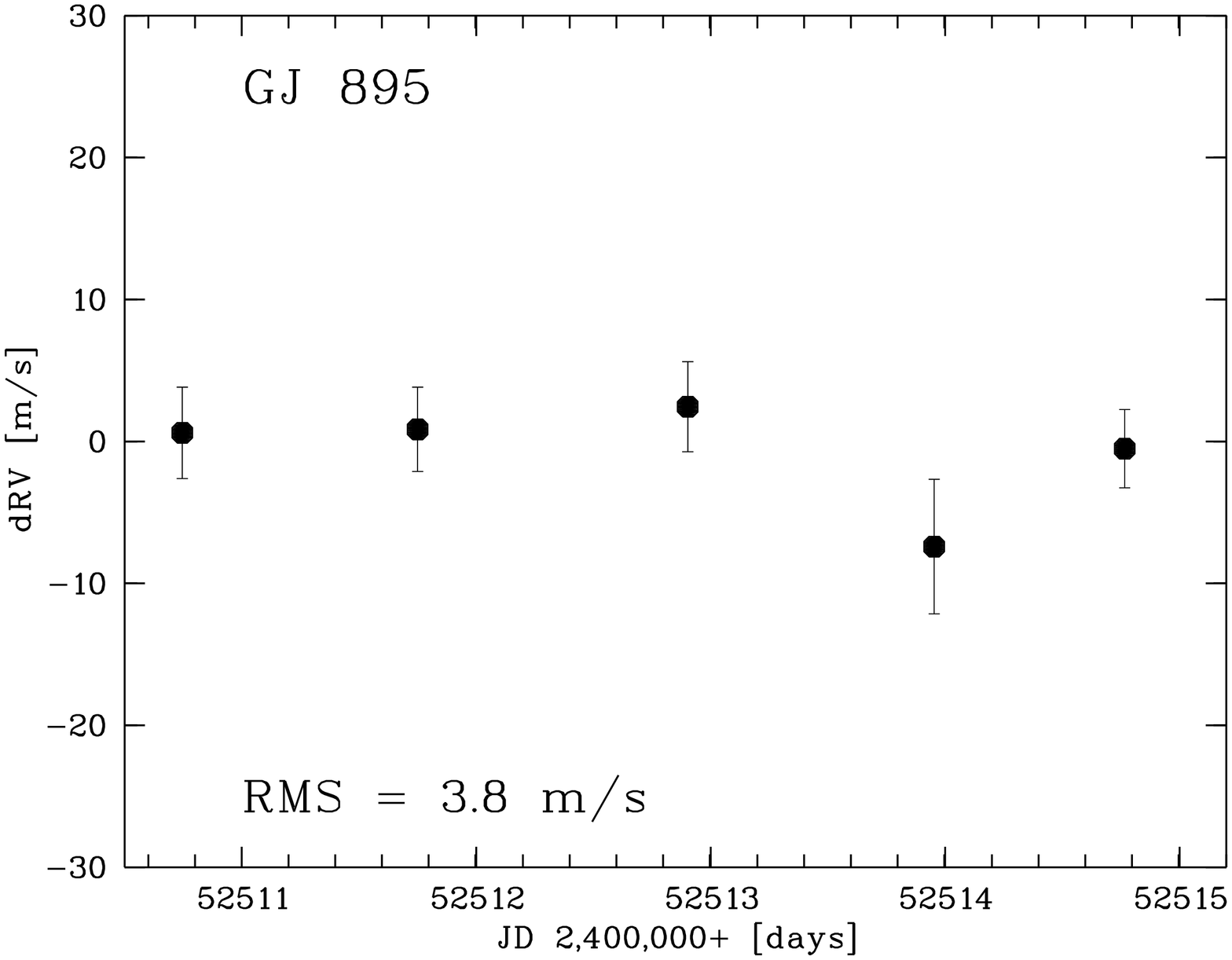,width=9.0cm,height=5.1cm,angle=270}}
        \vbox{\psfig{figure=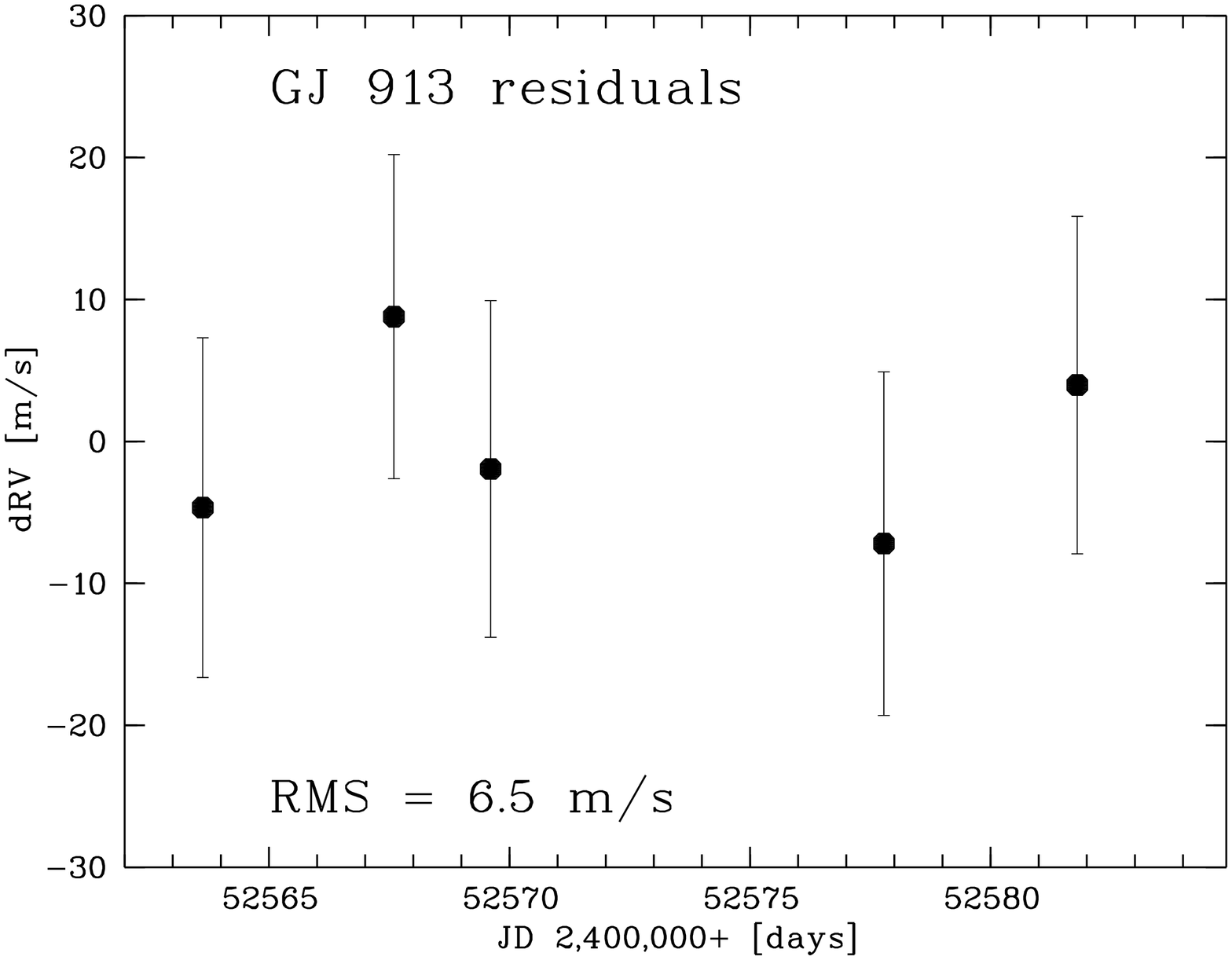,width=9.0cm,height=5.1cm,angle=270}}
   \par
        }
\caption[]{RV results for GJ~895 and GJ~913 (residuals).}
\label{rvsfig9}
\end{figure}

The RV-results of all targets listed in table 1 are displayed in Fig.~\ref{rvsfig1} to
Fig.~\ref{rvsfig9}. To allow a better comparison we plot all results on the same
velocity scale ($-30{~\rm m\,s}^{-1}$ to $+30{~\rm m\,s}^{-1}$), with the exception of
GJ~436 (Fig.~\ref{rvsfig5}) where the scale is enlarged. For GJ~623, GJ~748, GJ~864 and
GJ~913 we show the RV-data after subtraction of a large amplitude RV trend (see section 3.3).   

For one of the targets in the high-rms part of the distribution, GJ~436, we have already
obtained follow-up observations (see Fig.~\ref{rvsfig5}). A search for periodic
signals using the Lomb-Scargle periodogram (Lomb 1976, Scargle 1982) did not reveal any significant
signal in the power spectrum (the highest peak has a false-alarm-probability of $32\%$). 
The X-ray luminosity of $L_{X}=0.7 \times 10^{27}~{\rm erg\,s}^{-1}$ (H\"unsch et al. 1999) might point toward 
stellar activity as cause of the excess scatter. However, GJ~411 with a similar level of coronal emission
($L_{X}=0.6 \times 10^{27}~{\rm erg\,s}^{-1}$) and GJ~272 with $L_{X}=3.1\times 10^{27}~{\rm erg\,s}^{-1}$ 
the brightest X-ray source in our sample, do not show an increased RV-scatter. The cause of the 
variability of GJ~436 hence remains unknown. The other variable targets are currently in the 
observing queue of the HET and are pending re-observations.

\subsection{Linear and non-linear RV trends}

Three targets of our sample are either previously known (GJ~623, GJ~748) or suspected (GJ~913) binary stars.  

\begin{figure}
\centering{
  \vbox{\psfig{figure=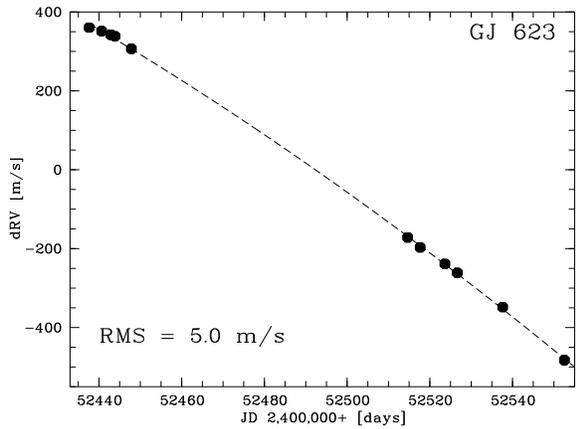,width=9.0cm,height=6.0cm,angle=270}}
   \par
        }
\caption[]{HET RV-data for the known binary GJ~623 showing a part of the binary orbit. The rms
scatter around the best-fit parabolic trend is $5.0~{\rm m\,s}^{-1}$ (see Fig.~\ref{rvsfig6} for the residuals).}
\label{gj623_curve}
\end{figure}

GJ~623 was found to be an astrometric binary by Lippincott \& Borgman~\cite{lippin} with
the latest published value for the orbital period of $3.7$ yrs (Nidever et al. 2002). Our
HET RV-results are shown in Fig.~\ref{gj623_curve} along with a parabolic fit to the data.  
Fig.~\ref{rvsfig6} displays 
the RV-residuals after subtraction of this trend. This data-set is now being used to further 
improve the binary orbital
solution and, in combination with HST FGS astrometry, to accurately determine the masses of the 
components of the GJ~623 system (Benedict priv.communication).
      
\begin{figure}
\centering{
  \vbox{\psfig{figure=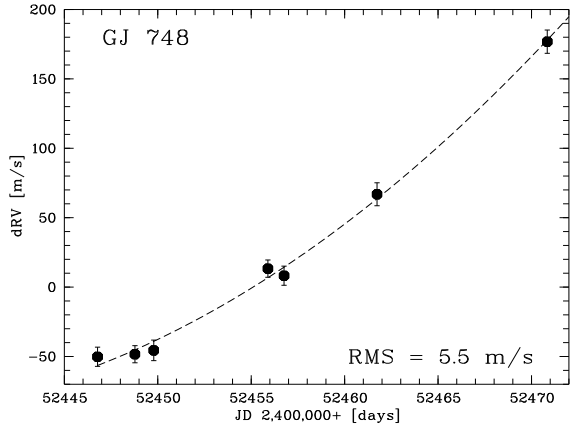,width=9.0cm,height=6.0cm,angle=270}}
   \par
        }
\caption[]{HET RV-data for the binary GJ~748 showing a small fraction of the binary orbit ($P=2.45$ yrs). 
The rms	scatter around the best-fit parabolic trend is $5.5~{\rm m\,s}^{-1}$ (see Fig.~\ref{rvsfig7} for the residuals).}
\label{gj748_curve}
\end{figure}

GJ~748 is another astrometric binary star (Harrington 1977), where we only see a small fraction
of the $P=2.45$ yrs (Franz et al. 1998) orbit (Fig.~\ref{gj748_curve}). The residual scatter around
a parabolic fit is $5.5~{\rm m\,s}^{-1}$ (Fig.~\ref{rvsfig8}).

\begin{figure}
\centering{
  \vbox{\psfig{figure=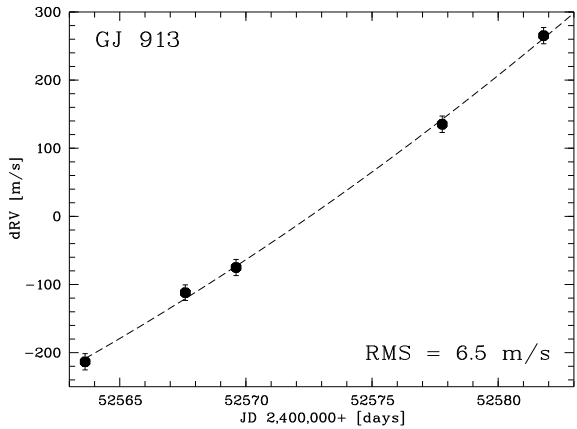,width=9.0cm,height=6.0cm,angle=270}}
   \par
        }
\caption[]{GJ~913, a Hipparcos candidate for a short-period binary system, shows a large
amplitude RV variation indicative of a stellar companion. The rms scatter around the best-fit parabolic trend is 
$6.5~{\rm m\,s}^{-1}$ (see Fig.~\ref{rvsfig8} for the residuals). Further observations might allow us
to determine a spectroscopic orbital solution.}
\label{gj913_curve}
\end{figure}

The entry of GJ~913 in the Hipparcos catalogue contains a double/multiple system annex flag X (stochastic model), 
which is a hint that the star is actually a short-periodic binary. 
The HET RV-data (Fig.~\ref{gj913_curve}) reveal a large amplitude
variation with a Dopplershift of $\approx 500~{\rm m\,s}^{-1}$ over the time span of $20$ days, 
indicative of binary orbital motion with an unknown period and thus confirming
the Hipparcos results. Further
observations will allow us to determine a spectroscopic orbit for this star.  

\begin{figure}
\centering{
  \vbox{\psfig{figure=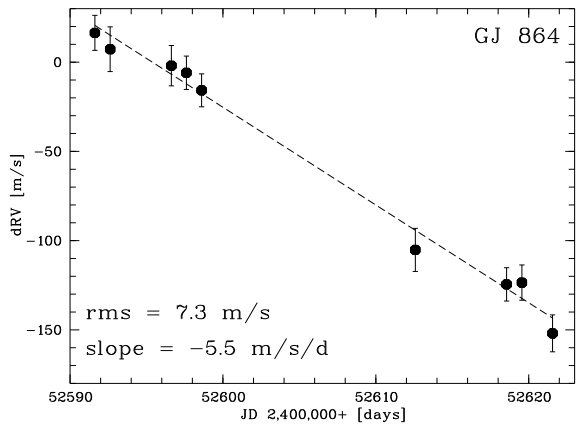,width=9.0cm,height=6.0cm,angle=270}}
   \par
        }
\caption[]{HET RV-data of GJ~864 showing a linear acceleration of $-5.5~{\rm m\,s}^{-1}{\rm d}^{-1}$.
The rms-scatter around this trend is $7.3~{\rm m\,s}^{-1}$. Using additional observations with the McDonald 2.7 meter 
telescope we confirmed that the linearity of this trend continued until January 2003 and that the best explanation is a 
previously unknown stellar companion at yet undetermined separation.}
\label{gj864_slope}
\end{figure}

In the case of GJ~864 we discovered a linear acceleration in the first $5$ RV measurements. Since
GJ~864 is not a known (or suspected) binary star and was classified as an RV-constant star
in the CORAVEL survey (Tokovinin~1992), we immediately re-scheduled GJ~864 for follow-up observations. 
As can be seen in Fig.~\ref{gj864_slope} the linear RV-trend continued until the star became unobservable 
by the HET in December 2002. Using auxiliary RV measurements with the McDonald 2.7 meter telescope in January
2003 we found that the RV-trend of GJ~864 remained linear without any sign of curvature and thus the
best explanation for the RV-data is a stellar companion at yet undetermined separation.      

\section{Discussion}

So far, we have found $5$ stars out of a sample of $39$ M-dwarfs which show a higher RV-scatter than
the rest of the sample and which we plan to monitor intensively in the near future to check if the
cause of the variability are short-periodic planetary companions. For GJ~436, the one target for which 
we already obtained follow-up observations, we could not find a periodic signal in the RV-data and the
cause of variability remains unknown.  

Two stars of our sample turned out to be previously unknown binary stars: GJ~864 and GJ~913. 
For both targets we soon might be able to establish a spectroscopic orbital solution. 

Despite their faintness M-dwarfs hold the promise to detect lower mass extrasolar planets than
what is usually possible by precise Doppler surveys. Due to the lower stellar primary mass even planets
with the mass of Neptune or lower become detectable in short-periodic orbits.     
For instance, a planet with $m\sin i = 17~{\rm M}_{\oplus}$ in a $4$ day orbit induces an RV semi-amplitude
of $\approx 11~{\rm ms}^{-1}$ on an M0V ($0.5~{\rm M}_{\odot}$) star, a $2\sigma$-signal for our
program. For later M-dwarfs these detection limits approach
the terrestrial planet regime as demonstrated by the mass-limits for Proxima Cen (Endl et al.~2003)
and Barnard's star (K\"urster et al.~2003). Thus, for the time being only precise Doppler monitoring of 
these low-mass stars has the capability
to detect planets with masses below $10~{\rm M}_{\oplus}$. Supposendly, such low-mass planets  
would have formed by planetesimal accretion, very similar to the way the Earth was formed, 
rather than by the processes which are responsible for the formation of gas giant planets (of a few 
hundred ${\rm M}_{\oplus}$). Clearly the detection of these kind of planets would be of special
interest for future space missions which will be dedicated to search for Earth analogs (KEPLER, NASA's TPF
\& ESA's Darwin).  
 
After the first year of the project the size of our target sample is still insufficient to arrive at any
significant statistical conclusions, but
we constantly add new targets to our list and the extension of our survey, both in target numbers as well 
as in time, will allow us in the
future to determine useful quantitative limits for planets around M-dwarf stars.

\acknowledgements
We are grateful to the McDonald Observatory TAC for generous allocation of observing time. 
The help and support of the HET staff and especially of the resident astronomers, Matthew
Shetrone, Brian Roman and Jeff Mader were also crucial for this project.    
We thank the referee for her/his helpful comments which improved the manuscript. 
We would also like to thank Sebastian G. Els for kindly providing his proper-motion routine 
for the velocity correction to the Solar System barycenter. Barbara McArthur and G. Fritz Benedict 
helped with many valuable discussions on M-dwarf binarity (moreover, Fritz helped with two sailing trips). 
We also thank G\"unter Wuchterl and Jack Lissauer for helpful discussions on planet
formation around M-dwarfs. 
This material is based upon work supported by
the National Aeronautics and Space Administration under Grant NAG5-9227 issued
through the Office of Space Science, and by National Science Foundation
Grant AST-9808980.

\begin{table}
\begin{center}
\begin{tabular}{l|lr|rrr}
\hline
\hline
Star & Sp.T. & V & N & rms & $\Delta$T \\
 & & [mag] & & [${\rm m\,s}^{-1}$] & [days] \\
\hline
GJ 2     & M2V & 9.93 & 5 & 2.6 & 20\\
GJ 87    & M2.5V & 10.06 & 5 & 8.8 & 40\\
GJ 155.1 & M1V & 11.04 & 5 & 6.1 & 16\\
GJ 184   & M0V & 9.93 & 5 & 7.3 & 41 \\
GJ 192   & M3.5V & 10.76 & 3 & 3.5 & 4\\
GJ 3352  & M3V & 11.07   & 6  & 5.0 & 23 \\
GJ 251.1 & M1.5V & 10.55 & 6 & 13.0 & 16\\
GJ 272   & M2V & 10.53   & 5 & 4.0 & 7 \\
GJ 277.1 & M0V & 10.49   & 6 & 8.7 & 14\\
GJ 281   & M0V & 9.61 & 4 & 6.6 & 15\\
GJ 289   & M2V & 11.46 & 5 & 7.6 & 52\\
GJ 308.1 & M0V & 10.33 & 8 & 12.4 & 40\\
GJ 310   & M1V & 9.30 & 7 & 15.9 & 71\\
GJ 328   & M1V & 9.99 & 5 & 6.2 & 42 \\
GJ 353   & M2V & 10.19 & 5 & 9.4 & 10\\
GJ 378   & M2V & 10.07 & 5 & 4.6 & 7\\
GJ 411   & M2Ve & 7.48 & 23 & 7.2 & 380\\
GJ 2085  & M1V & 11.18 & 5 & 6.2 & 10\\
GJ 436   & M3.5V & 10.67 & 17 & 20.6 & 394\\
GJ 535   & M0V & 9.03 & 5 & 3.0 & 27\\
GJ 552   & M2.5V & 10.68  & 6 & 11.3 & 33\\
GJ 563.1 & M2V & 9.71 & 6 & 4.9 & 24\\
GJ 623$^{*}$   & M3V & 10.28 & 11 & 5.0 & 116\\
GJ 671   & M3V & 11.37 & 7 & 6.1 & 12\\
GJ 687   & M3.5V & 9.18 & 5 & 2.9 & 10\\
GJ 699   & M4V & 9.53 & 16 & 7.2 & 36\\
GJ 708   & M1V & 10.07 & 6 & 9.2 & 23\\
GJ 4070  & M3V & 11.27 & 6 & 8.0 & 60\\ 
GJ 731   & M1.5V & 10.15 & 4 & 3.4 & 6 \\
GJ 748$^{*}$   & M4V & 11.10 & 7 & 5.5 & 26\\
GJ 839   & M1V & 10.35 & 6 & 6.3 & 14\\
GJ 849   & M3.5V & 10.37 & 4 & 1.2 & 34\\
GJ 864$^{*}$   & M1V & 10.01 & 9 & 7.3 & 31\\
GJ 895   & M2Ve & 10.04 & 5 & 3.8 & 5\\
GJ 913$^{*}$   & M0.5V & 9.62 & 5 & 6.5 & 19\\
\hline
\hline
\end{tabular}
\end{center}
\caption[]{
Radial velocity results for $35$ M-dwarfs ($4$ variable M-dwarfs of our program
are not listed here). Spectral classification and
visual magnitudes are taken from the Gliese catalogue, N is the total number
of obtained RV measurements, rms is the total scatter of the RVs and 
$\Delta$T denotes the duration of monitoring (i.e. the time span from 
first to last observation of this star).
Notes: for the $4$ stars marked with an asteriks the rms is given after subtraction 
of an RV-trend (see section 3.3).}
\label{all_stars}
\end{table}

\end{document}